**TITLE:**

# A CONTRIBUTION TO DISCERN THE TRUE IMPACT OF COVID-19 ON HUMAN MORTALITY

## AUTHOR

Stefano Barone, Ph.D.
Associate Professor of Statistics
University of Palermo, Department of Agricultural, Forest and Food Sciences, Palermo, Italy
stefano.barone@unipa.it
ORCID: 0000-0002-0339-8435

## ABSTRACT

The years 2020 and 2021 were characterized by the COVID-19 pandemic. The true impact of the pandemic on populations' health and life still has to be fully discerned. The main objective of this work is to discern the true impact of COVID-19 pandemic in the EU countries.

The mortality trends are considered and modelled. The excess mortality attributable to COVID-19, on a yearly basis, is estimated via a novel strategy that combines datasets from different official sources.

Considering demographic and geographic factors, new indices are also formulated, ranking the pandemic impact on EU countries, and new sociopolitical/economic reflections.

This work, which is also in line with the conclusions of previously published authoritative studies on excess mortality, represents an original methodology that, in a timely manner, can be implemented in the services of public decision makers for future emerging needs.

## COMPETING INTERESTS

No competing interests are involved in the research work and the presented results. No funding was supplied by any public/private agency or any other source for this research work.

# A contribution to discern the true impact of COVID-19 on human mortality

## ABSTRACT

The years 2020 and 2021 were characterized by the COVID-19 pandemic. The true impact of the pandemic on populations' health and life still has to be fully discerned. The main objective of this work is to discern the true impact of COVID-19 pandemic in the EU countries.

The mortality trends are considered and modelled. The excess mortality attributable to COVID-19, on a yearly basis, is estimated via a novel strategy that combines datasets from different official sources.

Considering demographic and geographic factors, new indices are also formulated, ranking the pandemic impact on EU countries, and new sociopolitical/economic reflections.

This work, which is also in line with the conclusions of previously published authoritative studies on excess mortality, represents an original methodology that, in a timely manner, can be implemented in the services of public decision makers for future emerging needs.

## 1 INTRODUCTION

The COVID-19 pandemic, caused by the severe acute respiratory syndrome coronavirus 2 (SARS-CoV-2), has significantly impacted humanity since its onset in 2020. However, in 2021, the pandemic was still ongoing, and vaccinations started. Understanding the real effect of the pandemic on mortality and the acceleration and amplification of demographic effects requires research efforts.

Studies have focused on the daily reported number of deaths related to COVID-19 and the clinical footprint to infer COVID-19 mortality. However, methods and metrics have been suggested, highlighting limitations concerning data collection and analysis.

Different countries used different protocols to test the most severe cases and report the eventual cause of death of a patient related to COVID-19. So, analysing COVID-19 mortality data in different countries has revealed large discrepancies.

To better discern the impact of COVID-19, an updated balance can be made by examining excess mortality data.

The excess mortality may be partly due to the epidemic disease and partly due to the excess burden caused by the disease within the healthcare system.

Recent studies involving many countries worldwide revealed that the excess mortality due to COVID-19 is much higher than the one officially reported, and highly variable between countries.

Ad hoc methods for modelling excess mortality data were set up and tested during the first wave of the pandemic. The World Health Organization (WHO) instituted a committee of experts to deepen the work on excess mortality, leading to impressive results starting in mid-2022.

This article presents a descriptive analysis of all-cause mortality data in EU countries for the years 2020 and 2021, highlighting the dramatic impact especially of the first two epidemic picks. The estimation of excess mortality through double exponential smoothing models for the all-cause mortality time series is adopted, and a novel strategy to better discern the true COVID-19 mortality is presented. Based on that, three new indices are also formulated, which allow ranking the pandemic impact on the EU countries and making sociopolitical/economic reflections.

The article is so structured: **Section 2** illustrates the methodology adopted in this research work. **Section 3** presents the results of the application of the methodology. **Section 4** provides a discussion and **Section 5** conclusions and future perspectives.

## 1.1 Literature review

*Zagheni (2021)* highlighted the importance of understanding the real effect of the COVID-19 pandemic on mortality and on the acceleration and amplification of demographic effects, which requires greater research efforts to be fully understood. Cross-national comparisons and reasoning on the effects of the pandemic, not only in terms of mortality but also in terms of fertility, has been also stressed (*Merchant 2021; Zhao 2021*).

Studies have focused on the daily reported number of deaths related to COVID-19 (*Millimet & Parmeter 2022*) and on the clinical footprint to infer COVID-19 mortality (*Rodríguez & Mena 2022*). Approaches and metrics were suggested, highlighting limitations concerning the data collection and analysis. Notably, different countries defined and used different protocols to test the most severe cases and to report the eventual cause of death of a patient related to COVID-19. However, even in the same country, such a major issue of death counting has occurred (*Cireraa et al. 2021*).

Analysis of COVID-19 mortality data in different countries has revealed discrepancies and variation in pandemic impact even in a limited geographical region (*Breen & Ermisch 2021*). Some studies explored the links between COVID-19 mortality and age (*Medford & Trias-Llimós 2020*).

The number of deaths was highly correlated with the number of cases and in the first pandemic wave the case fatality rate seemed to slowly converge toward a common value across countries (*Barone & Chakhunashvili, 2022*).

Pandemics developed in waves. Some countries experienced a higher mortality rate in the second wave than in the first wave; the opposite occurred in other countries (*Morwinsky, Nitsche & Acosta 2021*).

All studies, although methodologically sound, lose efficacy if the data on mortality due to COVID-19 are not reliable.

With the accumulated data and acquired knowledge, and owing to the abovementioned problems of accounting, after the first — and the most dramatic — two years of the pandemic, an updated balance could also be made by examining *excess mortality*.

Studies on excess mortality with established methods and models are not new (see, e.g., *Nepomuceno et al. 2022* and references therein) or epidemiology (see, e.g., *Fouillet et al. 2006; Housworth & Langmuir 1974*).

A research question that has interested several researchers is to explore the excess mortality data of 2020 and 2021, to better discern the impact of COVID-19.

For example, in November 2020 in Italy, 76291 officially reported deaths occurred (data source: *ISTAT 2022*). In the previous five years (2015--2019), the average number of deaths in the same month was 51462. Therefore, in November 2020, there were 24829 more deaths than the average of the previous five years. However, only 16583 deaths were officially reported as being due to COVID-19 (data source: *ECDC 2022*).

The dramatic impact of the pandemic in Italy in 2020 in terms of mortality has already been well documented (see e.g. *Rovetta & Bhagavathula 2022*). Also in France, considerable excess mortality occurred in the period March–April 2020 (*Dahoo & Gaudy, 2020*).

Comorbidities and institutional factors can be covariates in the analysis of COVID-19 mortality (*Rodrıguez & Mena 2022*; *Najera & Ortega-Avila 2021*). Some authors have calculated the excess mortality due to COVID-19 on small islands, noting that small, isolated communities are characterized by lower mortality rates (*Riccò et al. 2022*). Researchers found that many covariates may have had either positive or negative effects on mortality during the pandemic; however, their conjoint effect is currently considered neutral.

Rapid analysis of mortality during an epidemic is a priority for surveillance and control, and the excess mortality may be partly due to the epidemic disease and partly due to the excess burden caused by the disease within the health care system (*Setel et al. 2020*).

In densely populated and complex countries, such as Brazil, the use of excess mortality to assess the impact of the pandemic and for epidemiological surveillance was promptly used (*Freitas et al. 2020*).

The comparison between excess mortality and officially reported COVID-19-related mortality was performed with data from many countries worldwide by *Achilleos et al.* (*2022*).

Additionally, in Iran (*Safavi-Naini et al. 2022*) and Russia (*Kobak 2021*), a large discrepancy between excess mortality and COVID-19-related mortality was observed, suggesting the adoption of excess mortality as an indicator of the true impact of the pandemic.

Recent studies (*Karlinsky & Kobak 2021*; *Wang et al. 2022; Msemburi et al. 2022*) involving many countries worldwide revealed that the excess mortality due to COVID-19 is much higher than the one officially reported, and highly variable between countries.

A recent article (*Śleszyński et al. 2023*) analyzing mortality data in Europe on a detailed spatial scale, has also revealed that the SARS-CoV-2 virus might have had fatal effects largely before 2020.

According to excess mortality analysis, COVID-19 was the third leading cause of death in 2020 in the USA (*Fricker 2021*). Still concerning USA, *Paglino et al.* (*2023*) use a Bayesian hierarchical model to estimate the excess mortality due to COVID-19.

Statistical methods for modelling excess mortality data were set up and tested during the first wave of the pandemic in Italy (*Henry et al. 2022*).

The first evidence of the use of excess mortality to assess the impact of the COVID-19 pandemic, together with the evidence of deficiencies in data collection of COVID-19 cases and related deaths, convinced the World Health Organization (WHO) to institute a committee of experts to deepen the work on excess mortality, which led to impressive results starting in mid-2022 ("*Estimates of Excess Mortality…" 2022*, *WHO online document 2023*, *Knutson et al. 2022*).

*Nepomuceno et al. (2022)* present a brilliant work in which they compare the use of several mortality indices, expected mortality models, time series lengths, and different time scales of data to evaluate the effect on the estimation of excess mortality. However, they don't find a universally best method to estimate the excess mortality.

(*Paglino & Stokes* (*2024*) note that excess mortality analysis requires relatively little data, does not depend on accurately assigned causes of death, and can be compared across space and time.

The work presented here follows the conclusions of most of the above-mentioned references.

# 2 METHODOLOGY

## 2.1 Definitions and abbreviations.

Revising the wide and cross-disciplinary literature on the topic, the following definitions and abbreviations are adopted throughout the paper:

| ***Name*** | **Abbreviation** | **Definition** |
|---|---|---|
| *Actual Deaths* | AD | Number of deaths declared by official sources (e.g. EUROSTAT) |
| *All-cause mortality* | - | Number of deaths occurred for any cause |
| *Case fatality rate* | - | Number of deaths due to a disease/number of people affected by the disease |
| *Country code* | - | Numeric code assigned to each of the EU countries (in alphabetic order) |
| *COVID Cases* | CC | Number of COVID infections declared by official sources (e.g. ECDC) |
| *COVID-19 fatality rate* | $I'_F$ | See Methodology Section |
| *COVID-19 mortality* | - | Number of deaths attributed to COVID-19 |
| *Deaths by COVID* | DC | In the article they are: officially declared DC, predicted by CC or predicted by EM |
| *Discrepancy Index* | $I_D$ | See Methodology Section |
| *European Union* | EU | |
| *Excess Mortality* | EM | Number of deaths excessive respect to the one which could be forecasted by the previous year in a specific region (e.g. a country) |
| *Fatal Impact Index* | $I_F$ | See Methodology Section |
| *Forecasted all-cause mortality* | $\widehat{D}$ | Expected deaths obtained by modelling the time series and applying a forecasting procedure for one-two steps forward |
| *Land Area* | A | The area [$km^2$] of the surface covered by a region (country) |
| *Lower Bound* | LB | For a confidence interval |
| *Mean predicted DC* | $\overline{\mathrm{DC}}$ | The average between the DC values predicted by CC and by EM |
| *Mortality rate* | - | Number of deaths/population |
| *Population* | P | Number of inhabitants in a specific region (e.g. a country) |
| *Upper Bound* | UB | For a confidence interval |

The methodology adopted in this work can be outlined through the following phases, which will be illustrated in this Section:

1. Modelling the all-cause mortality time series
2. Forecasting all-cause mortality in 2020 and 2021 through the time series 2010-2019.
3. Deriving the excess mortality (EM) for 2020 and 2021 as difference $(\mathrm{AD} - \widehat{D})$
4. Using EM and CC to better estimating DC
5. Formulating discrepancy and fatality indices for the EU countries for 2020 and 2021.

### 2.2 All-cause mortality trends in the EU countries

Owing to problems concerning the classification of COVID-19-related deaths, researchers have analysed data concerning all-cause mortality.

The research question about how mortality 2020--2021 does relate to trends in previous years should be extended to a sample of countries large enough to allow general considerations, but not too large, to avoid problems of heterogeneous populations. Therefore, in this work, the analysis was limited to the 27 EU countries. All-cause mortality data were downloaded from the official database EUROSTAT (2022) and thoroughly analysed.

### 2.3 Estimation of the excess all-cause mortality for 2020 and 2021

The time series of all-cause mortality for EU countries moves the reasoning to excess mortality. Excess mortality is systematically studied and officially reported by the EU agency EUROMOMO. The excess mortality over a time frame is the difference between the number of deaths that occurred and what was expected (so-called "mortality baseline") in turn estimated from the past. They adopt a Poisson g.l.m., corrected for overdispersion (EUROMOMO 2022, "Methods" section, and reported references). The rates of excess mortality throughout Europe, which increased until the end of 2020 and increased until the end of 2021, were very close to each other, and both were very different from those in the previous three years. This confirms that 2021 was characterized by the presence of a special cause of mortality still dramatically weighing the European population.

The analysis of excess all-cause mortality, restricted to the case of the in Europe, differed. The cumulative excess mortality in 2021, which was lower than that in 2020 in the first nine months of the year, tended to increase at the end of the year. Still for the 0–14 year age group, the 2021 trend was in line with the (prepandemic) past.

In this work, the idea is to split the time series 2010--2021 into two periods: the prepandemic period 2010--2019 and the period 2020--2021. A *double exponential smoothing* (also called Holt's exponential smoothing) is here adopted to model the mortality time series in the prepandemic period and to forecast the years 2020 and 2021

The model is made of two components, *level* and *trend*, at each time point (*year* in this case). The model uses two *smoothing constants*, α and $\gamma$ ($0 \leq \alpha, \gamma \leq 1$), to update the two components at each time point.

The updating equations are as follows:

$$L_t = \alpha \cdot Y_t + (1 - \alpha) \cdot [L_{t-1} + T_{t-1}] \qquad (1)$$

$$T_t = \gamma \, [L_t - L_{t-1}] + (1 - \gamma) \, T_{t-1} \qquad (2)$$

where:

$L_t$ is the *level* at time $t$

$T_t$ is the *trend* at time $t$

$Y_t$ is the (mortality) data value at time $t$

Equation (1) indicates that the level of the data at time point *t* is partially explained by the registered value *Y* at that time and partially explained by the forecasting (level + slope) from the previous time point.

Equation (2) indicates that the trend (slope) of the data series at time point *t* can be partially explained by the gap in levels between $t$ and $t-1$ and partially explained by the trend estimated at the previous time point.

The initial values for the level and trend ($L_0$ ; $T_0$) and the smoothing constants are estimated through an *optimal ARIMA*(0,2,2) model fitted to the 2010--2019 data.

The forecasts for time points $t+1$ and $t+2$ are given by:

$$\hat{Y}_{t+1} = L_t + T_t \qquad (3)$$

$$\hat{Y}_{t+2} = L_t + 2 \, T_t \qquad (4)$$

The prediction limits (lower bound LB, upper bound UB) are based on the *mean absolute deviation*

(MAD) accuracy measure; the formulas are given in *Farnum & Stanton (1989)*.

As a validation of the method, the double-exponential smoothing model has been used to compare – for each EU country – the actual mortality values of 2018 and 2019 with the forecasted ones obtained by using the 2010—2017 time series.

### 2.4 Using the excess all-cause mortality to better estimate COVID mortality

The estimated excess mortality, obtained by difference between the actual all-cause deaths and the forecasted all-cause deaths, can be utilized to better estimate COVID-19 mortality.

In fact, the estimated excess mortality (henceforth abbreviated EM), "covid cases" (abbreviated CC) and official deaths "by covid" (abbreviated DC) are correlated and independently derived.

From data driven reasoning (the reader is invited to refer to the section Results) it can be seen that:

- the strongest correlations concern DC-CC and DC-EM;
- simple linear regression models could not be adopted. In fact, an inflated variation occurring for increasing values of “x” and “y” variables and an increase in variation for 2020--2021 makes the linear model inappropriate for the original variables, suggesting a *log* transformation;
- The regression models on the log-log scales fit data very well.

By using the two regression models, it is possible to predict DC in two different ways: for a given CC and for a given EM. Therefore, separately for each country, it is possible to derive two values of the predicted DC, which can be compared to the officially declared value of the DC.

This strategy, schematically depicted in Figure 6, stems from the reasoning that there is less uncertainty in CC and EM data than in the officially declared DC.

### 2.5 Discrepancy and fatality indices

During the most dramatic years of the pandemic, i.e., 2020 and 2021, there have been continuous and exhausting debates on the true numbers of deaths due to COVID-19 compared with those officially declared by various countries.

With the present methodology, which is able to estimate COVID-19 mortality data with greater precision, it is therefore possible to determine a *discrepancy index*, which can be useful for making critical reasoning and comparisons between the EU countries.

To define a discrepancy index, it is first necessary to *standardize* the number of deaths, dividing it by a "reference" population.

Let us define:

$d_1$ = standardized min predicted DC

$d_2$ = standardized max predicted DC

$d_3$ = standardized officially declared DC.

The minimum and maximum predicted DC values are the smallest and the largest of the two DC values predicted by CC and EM, respectively. The values of $d_1$, $d_2$ and $d_3$ can be easily calculated.

However, a complete comparison should be made by considering at least two aspects:

i) The distance $d_2 - d_1$, i.e., the distance between the maximum and the minimum standardized predicted DC, for such a distance (meaning prediction uncertainty) the ideal is "the lower the better".

ii) The absolute value of the distance between the standardized declared DC and the mean value of the two standardized predicted DC. The lower this distance (meaning that the declared DC is close to the "most realistic value") is, the better.

On the basis of previous considerations, a ***discrepancy index*** $\boldsymbol{I_D}$ is formulated as follows:

$$I_D = (d_2 - d_1) + \left| d_3 - \frac{1}{2}(d_1 + d_2) \right| \qquad (5)$$

The discrepancy index considers the previous aspects *i)* and *ii)*, giving them the same importance.

From the previous analyses, it is assumed that a measure of the pandemic impact can be the mean predicted DC (henceforth denoted $\overline{DC}$), which makes further considerations possible.

Certainly, the absolute value of the number of deaths is meaningful. However, the relative weight should also be considered. The number of deaths should be referred to the corresponding population, since there is no doubt that one thing is to suffer ten deaths over a population of one hundred subjects, another is to suffer ten deaths over a population of one thousand subjects, with the first case intuitively being worse than the second.

Moreover, mortality should also be considered in relation to a particular geographical area, since one thing is to suffer the loss of ten individuals over a territory that is a hundred square kilometres; another thing is to refer to a territory of one thousand square kilometres. The second case is intuitively less critical.

Therefore, with data of $\overline{DC}$, population (*P*), and land area (*A*) for the EU countries (see Table 4), a ***pandemic fatal impact index $I_F$*** can be calculated:

$$I_F = \overline{DC}/(P \cdot A) \qquad (6)$$

In this type of analysis (index $I_F$), the small countries are penalized. However, this is the reality, unless an absurd reasoning is made so that the higher the population density is, the greater the number of losses the better. In this case, land area *A* would be placed in the nominator of the ratio (6).

Hence, to mitigate the effect of land area and to complete the framework of these comparative analyses, by removing *A* from (6), another index is derived, which can be defined as ***COVID-19 fatality rate***, given by the ratio of the mean predicted DC and the population size:

$$I'_F = \overline{DC}/P \qquad (7)$$

# 3 RESULTS

## 3.1 All-cause mortality trends

An exploratory analysis, which can be taken as a starting point, concerns the case of Italy, which has been considered the most impacted country in the EU, especially during the COVID-19 pandemic outbreak period.

For this analysis, data were downloaded from the Italian Institute of Statistics (*ISTAT 2022*). Data concerning all-cause mortality in Italy can be considered unaffected by uncertainty. The mortality data for all causes in the Italian territory from 2020--2021 and for the previous five-year period from 2015--2019 revealed that in 2020, there were two major peaks in mortality compared with the past: in the periods from March--April and October–December. Additionally, in the other months, the mortality rate seemed to be higher than that in the past. Overall, in the twelve months of 2020, there were more than 100,000 deaths than the average of the previous five years. Concerning 2021, two peaks in mortality could still be observed in the periods March–April and October–December but were lower than those in 2020 (note that these two periods in 2021 were affected by lockdowns, i.e., tight restrictions to people's circulation). For the rest of the year, the overall mortality did not differ much from that in 2020 and was even higher in some months.

For the 27 EU countries, all-cause mortality data, starting from the year 2010, downloaded from the official database EUROSTAT (2022), are visually summarised on Figure 1.

Stationary trends with a spike in 2020/2021 can be observed only in a few countries, while in most countries, the mortality trend was already increasing. Except for Ireland, which shows a steady increasing trend, the other trends show a sharp acceleration in 2020. A deceleration in 2021 appears only in some cases (visibly, Belgium, France, Italy, Luxembourg, Slovenia, Spain, Sweden).

## 3.2 Estimation of the excess all-cause mortality for 2020 and 2021 in the EU Countries

The calculations and forecasts of all-cause mortality for 2020 and 2021, for the 27 EU countries were performed in MINITAB©. Figure 2 shows an example of output (country: Austria). The original time series of deaths, the smoothed series, and the forecasts of the years 2020 and 2021 (predicted values and confidence intervals) are plotted in the upper left side of the figure. The lower left side of the figure shows the analyses of the residuals, which were satisfactory for all 27 EU countries.

All the forecasted values of mortality in 2020 and 2021, which are based on trends 2010—2019, and calculated through the adopted method, are reported in Table 1 (together with the actual deaths provided by EUROSTAT). Note that the actual deaths registered in 2020 and 2021 lie outside the prediction confidence intervals in most of the cases.

The excess all-cause deaths (indicated by EM in Table 1) are calculated by difference between the actual all-cause deaths and the forecasted all-cause deaths.

Table 2 provides a summary of the method validation. It is very satisfactory: the Mean Absolute Percentage Error (MAPE) for the prediction of 2018 is equal to 1.5% for the prediction of 2019 MAPE=2.1%.

### 3.3 Estimation of COVID mortality

The last six columns of Table 1 provide the estimated excess mortality (EM), "covid cases" (CC) and official deaths "by covid" (DC) for 2020 and 2021.

By plotting the data in 3D scatterplots (for both 2020 and 2021), a clear correlation is evident between the EM, CC, and DC (Figure 3). Notably, the three variables are independently derived. Moreover, by projecting the data on the three coordinate planes, it is possible verify that the strongest correlations concerned DC-CC and DC-EM and simple linear regression models cannot be adopted.

A summary of the regression analyses of DC vs. CC and DC vs. EM for the year 2020 is shown in Figure 4. Note that the regression analysis DC-EM for the year 2020 is performed on 26 data points instead of 27, excluding Denmark, since a negative predicted EM made the log transformation impossible.

The regression models fit the log-log scales well.

By using the two regression models, it is possible to predict DC for a given CC and for a given EM. Therefore, separately for each country, it is possible to derive two values of the predicted DC, which can be compared to the officially declared value of the DC.

The overall results are reported in Table 3 for both 2020 and 2021, whereas (b)

Figure 7a (only for 2020) plots the values of Table 3 on a log scale to facilitate comparisons between the countries. (b)

Figure 7a shows that the DC predicted by the CC (orange dash symbol) is sometimes higher and sometimes lower than the DC predicted by the EM (blue dash symbol). The numbers of the two cases are almost the same (14 vs. 13). This is an outstanding result, meaning that the two predictive methods are balanced.

The officially declared DC (black cross symbol) lies within the two predicted values only in some cases, whereas in most cases, it lies over or below the prediction interval. This implies that the declared DC can be either considered *reliable* or not.

### 3.4 Discrepancy and fatality indices

To calculate the discrepancy index, the number of deaths are *standardized* by using reference population numbers (Table 4, second column, P).

Table 4 reports the calculated discrepancy index $I_D$ for 2020 and 2021. Figure 8 plots the discrepancy index in bar charts (sorted from high to low), showing that the worst cases (Spain and Romania) largely exceed all others in 2020 and 2021.

The bar chart of (b)

Figure 9.a shows the calculated *Fatal Impact Index* $I_F$ (Table 4) sorted from the highest to the lowest. According to this chart, the interpretation of the pandemic's fatal impact radically changes with respect to the common opinion, indicating that the small country of Malta was impressively

impacted, much farther from other EU countries.

By removing the cases of the very small countries Cyprus, Luxembourg and Malta, which cause crushing of the other bars, the relative fatal impact on the other countries becomes more visible ((b) Figure 9.b).

A note should be reserved for Sweden, since its "land area" includes a vast inhabited area due to severe climatic reasons. If realistically considering 50% of the area, Sweden would lie between Estonia and Italy in 2020 and between Italy and Poland in 2021, making it slightly more critical.

The bar charts in Figure 10 show the calculated *Fatality Rates* $I'_F$ values sorted from the highest to the lowest values for 2020 and 2021. These plots depict a more balanced situation between the EU countries but a fatality ranking that changes substantially from 2020--2021 and provides a picture of the pandemic impact rather distant from the common opinion.

# 4 DISCUSSION

Italy, the most impacted country in the EU, experienced two major peaks in mortality during the COVID-19 pandemic, with over 100,000 deaths in the twelve months of 2020. The mortality rate seemed to be higher than previous years, with two peaks observed in March-April and October-December. In 2021, two peaks were observed in March-April and October-December but were lower than in 2020 due to lockdowns. Overall mortality did not significantly differ from 2020 and was even higher in some months.

For the 27 EU countries, all-cause mortality data from 2010 were visually summarized. Stationary trends with a spike in 2020/2021 can be observed in a few countries, while most countries show an increasing mortality trend. Ireland showed a steady increasing trend, while other trends showed a sharp acceleration in 2020. A deceleration in 2021 appears only in some cases, such as Belgium, France, Italy, Luxembourg, Slovenia, Spain, and Sweden.

The time series were modelled through double-exponential smoothing, a well-established method for time series showing trends. The modelling, applied to the 2010—2019 data, allowed forecasting of all-cause mortality for 2020 and 2021 in the 27 EU countries. The previous Section provides examples of the original time series of deaths, the smoothed series, and the forecasts. The residuals analyses are satisfactory. The method was validated, with a Mean Absolute Percentage Error of 1.5% for 2018 and 2.1% for 2019.

Table 1 reports all forecasted mortality values, along with official deaths provided by EUROSTAT. The excess all-cause deaths were calculated by difference between the actual and forecasted all-cause deaths.

Table 1 presents estimated excess mortality (EM), covid cases (CC), and official deaths (DC) for 2020 and 2021. 3D scatterplots reveal a clear correlation between the variables, but simple linear regression models cannot be adopted. The regression analyses of DC vs. CC and DC vs. EM for 2020 are illustrated in Figure 4 for one case. The regression models fit log-log scales well, allowing for the prediction of DC for a given CC and EM. This allows for two values of the predicted DC to be derived for each country, which can be compared to the officially declared DC.

Overall results are reported in Table 3 for both 2020 and 2021, with Figure 6a plotting the values on a log scale for 2020. The DC predicted by the CC is sometimes higher or lower than the DC predicted by the EM, indicating that the two predictive methods are balanced. The officially declared DC lies within the predicted values in some cases, while in most cases, it lies over or below the prediction interval, indicating the unreliability of the declared DC.

The *Discrepancy Indices*, calculated by standardizing the number of deaths using reference population numbers shows that the worst cases, Spain and Romania, largely overshadow the other EU countries both in 2020 and 2021. The *Fatal Impact Indices* show that Malta was significantly impacted. However, apart from Cyprus, Luxembourg, and Malta cases, the *Fatal Impact Index* makes more visible a ranking of the impact in the EU countries. Sweden, with its vast inhabited area should be evaluated more carefully. The *COVID-19 Fatality Rates* values for 2020 and 2021 show a more balanced situation between EU countries, but, impressively, the fatality ranking changes substantially from 2020 to 2021, providing a picture of the pandemic impact that is distant from common opinion.

Concerning the last two indices, which are obtaining dividing the estimated COVID-19 Excess Mortality by a reference population, it has to be noted that the age structure of the populations were not taken into account, partly colliding with previous studies (*Nepomuceno et al. 2022; Paglino & Stokes 2024*). However, it has to be considered that the focus here was placed on estimating excess mortality only for 2020 and 2021, which was mostly due to COVID-19. Since it has also been ascertained that the disease was primarily lethal for adults, especially the elderly, age structure should have a marginal impact; moreover, the age structure is not so variable between the 27 EU countries (see e.g. the median age comparison on EUROSTAT website).

Anyhow, this can be still considered a limitation of this work, requiring further insights,

# 5 CONCLUSIONS

Remarkable excess mortality occurred in 2020 worldwide, which is currently clearly attributed to the COVID-19 pandemic. At the EU level, the excess mortality in 2020 due to COVID-19 can be here estimated as lying between 308,353 and 367,291 deaths (totals of the 3rd and 4th columns of Table 3), vs. an official declared number of deaths of 373,266 deaths (total of the 5th column of Table 3). The year 2020 was the year of the pandemic outbreak, with rather free circulation of the new virus, an absence of vaccines and adequate treatments. For completeness of reasoning and to be consistent with previous studies, the year 2021 was included in this analysis, as it was characterized by better knowledge of the virus, vaccinations and more adequate treatments.

In terms of the excess mortality for 2021, compared with 2020 and the past, unfortunately, we do not see any improvement. Excess mortality due to COVID-19 in the EU in 2021 can be here estimated as lying between 394,768 and 502,073 deaths (totals of the 6th and 7th columns of Table 3), vs. an official declared number of deaths of 525,909 deaths (total of the 8th column of Table 3). Therefore, it is undoubtedly higher than that of 2020 and affected by higher uncertainty (the spread between the lower and the upped *bound* is double respect to 2020).

The estimates of excess mortality are based on predictive models that consider the previous history. Therefore, over time, if the years 2020 and 2021, characterized by such a jump in mortality, are considered, the baseline will be inflated so that we will be increasingly less aware of the dramatic effect of COVID-19 in terms of mortality.

This work was based on accurate statistical analysis and the use of inferential methods with publicly available datasets. The aim is to reach a wide readership, as the topic is globally important and the involved issues still need clarification and better solutions for the future. It was given importance to clear tables and graphics.

This analysis referred to EU countries where the management of the pandemic was rather homogeneous due to common EU directives. Conversely, in other countries, as China, South Korea and Australia, pandemic management has been deployed in completely different ways.

This work, in line with the conclusions of previously published papers on demographic studies of excess mortality, represents an original methodology that – regularly implemented – can constitute a tool at the service of public decision makers for future emerging needs.

## DATA AVAILABILITY

All necessary data for the analyses presented in this article were extracted and downloaded from the official websites of ISTAT (2022), ECDC (2022), EUROSTAT (2022) and EUROMOMO (2022). Computations were performed via MINITAB and MS Excel. Datasets and calculation files are available on a public web repository.

Population Council, 2021.

**Table 1. Actual and predicted deaths; Estimated Excess Mortality, Covid Cases, and Deaths by Covid for 2020 and 2021 for the 27 EU countries.**

| | | | Prediction 2020 | | | | Prediction 2021 | | | | | | | | |
|---|---|---|---|---|---|---|---|---|---|---|---|---|---|---|---|
| **Country** | Country code | AD2020 | LB | $\widehat{D}$ | UB | AD2021 | LB | $\widehat{D}$ | UB | Estimated EM (2020) | CC (2020) | DC (2020) | Estimated EM (2021) | CC (2021) | DC (2021) |
| Austria | 1 | 91599 | 81943 | 85153 | 88364 | 91962 | 82443 | 85938 | 89433 | 6446 | 360767 | 6431 | 6024 | 914787 | 9197 |
| Belgium | 2 | 126896 | 106409 | 111057 | 115706 | 112331 | 106245 | 111455 | 116664 | 15839 | 650041 | 19819 | 876 | 1475862 | 8563 |
| Bulgaria | 3 | 124735 | 104997 | 108823 | 112648 | 148995 | 104781 | 108903 | 113026 | 15912 | 202266 | 7515 | 40092 | 544842 | 23375 |
| Croatia | 4 | 57023 | 50289 | 52900 | 55510 | 62712 | 50210 | 53006 | 55801 | 4124 | 210837 | 3920 | 9707 | 504408 | 8618 |
| Cyprus | 5 | 6422 | 5544 | 6174 | 6805 | 7110 | 5616 | 6296 | 6977 | 248 | 15101 | 78 | 814 | 144176 | 568 |
| Czechia | 6 | 129289 | 109039 | 113359 | 117680 | 139891 | 109614 | 114245 | 118877 | 15930 | 732567 | 12016 | 25646 | 1752631 | 24339 |
| Denmark | 7 | 54645 | 53966 | 55835 | 57703 | 57152 | - | 56216 | - | -1190 | 156508 | 907 | 936 | 606526 | 1938 |
| Estonia | 8 | 15811 | 15345 | 15775 | 16206 | 18587 | - | 15738 | - | 36 | 28393 | 233 | 2849 | 213738 | 1692 |
| Finland | 9 | 55488 | 53986 | 54952 | 55918 | 57659 | 54342 | 55369 | 56396 | 536 | 36676 | 598 | 2290 | 241354 | 1150 |
| France | 10 | 669137 | 607893 | 627025 | 646157 | 657134 | 613220 | 635195 | 657170 | 42112 | 2620425 | 44826 | 21939 | 7352375 | 59135 |
| Germany | 11 | 985572 | 932157 | 966461 | 1000765 | 1023687 | 940102 | 978323 | 1016545 | 19111 | 1754283 | 51010 | 45364 | 5440699 | 67134 |
| Greece | 12 | 130620 | 120292 | 126519 | 132747 | 143329 | 121322 | 128120 | 134918 | 4101 | 138850 | 4838 | 15209 | 1072003 | 20790 |
| Hungary | 13 | 141326 | 126519 | 131167 | 135814 | 156131 | 126577 | 131634 | 136692 | 10159 | 322514 | 9537 | 24497 | 933901 | 29649 |
| Ireland | 14 | 31765 | 30834 | 31482 | 32130 | 33058 | 31053 | 31753 | 32454 | 283 | 91779 | 2239 | 1305 | 696780 | 3675 |
| Italy | 15 | 746146 | 612396 | 648310 | 684225 | 709035 | 613760 | 652604 | 691447 | 97836 | 2107166 | 74159 | 56431 | 4018517 | 63243 |
| Latvia | 16 | 28854 | 27185 | 28166 | 29147 | 34600 | 27017 | 28071 | 29125 | 688 | 40904 | 635 | 6529 | 235770 | 3935 |
| Lithuania | 17 | 43547 | 36376 | 38322 | 40269 | 47746 | 35562 | 37879 | 40196 | 5225 | 147990 | 1845 | 9867 | 378766 | 5571 |
| Luxembourg | 18 | 4609 | 4196 | 4404 | 4611 | 4489 | 4237 | 4482 | 4727 | 205 | 41272 | 396 | 7 | 58282 | 407 |
| Malta | 19 | 4084 | 3492 | 3755 | 4017 | 4163 | 3545 | 3825 | 4106 | 329 | 11101 | 166 | 338 | 39696 | 258 |
| Netherlands | 20 | 168678 | 151155 | 155760 | 160366 | 170972 | 152799 | 157786 | 162773 | 12918 | 795539 | 11405 | 13186 | 2336036 | 9514 |
| Poland | 21 | 477355 | 399460 | 415932 | 432405 | 519517 | 403476 | 421148 | 438819 | 61423 | 1294878 | 28554 | 98369 | 2813337 | 68500 |
| Portugal | 22 | 123357 | 109069 | 113315 | 117561 | 124802 | 109741 | 114374 | 119006 | 10042 | 413678 | 6906 | 10428 | 1013820 | 11989 |
| Romania | 23 | 297039 | 255712 | 264713 | 273713 | 334473 | 256484 | 266069 | 275654 | 32326 | 632263 | 15767 | 68404 | 1176628 | 42985 |
| Slovakia | 24 | 59089 | 51964 | 54054 | 56144 | 73461 | 52075 | 54318 | 56561 | 5035 | 285091 | 2250 | 19143 | 1090159 | 14415 |
| Slovenia | 25 | 24016 | 20287 | 20942 | 21597 | 23261 | 20492 | 21198 | 21905 | 3074 | 124034 | 3025 | 2063 | 341757 | 2885 |
| Spain | 26 | 491602 | 411958 | 430694 | 449429 | 449270 | 414323 | 434687 | 455052 | 60908 | 2009358 | 54374 | 14583 | 4813486 | 36850 |
| Sweden | 27 | 98124 | 88166 | 90981 | 93796 | 91958 | 87874 | 90930 | 93987 | 7143 | 454758 | 9817 | 1028 | 881840 | 5534 |

*Notes*: AD=Actual Deaths (source: EUROSTAT); LB=Lower Bound; UB=Upper Bound; $\widehat{D}$=Forecasted Deaths; EM=Excess Mortality; CC = Covid Cases; DC = Deaths by Covid

**Table 2. Validation of the double exponential smoothing model.**

| **Country** | Country code | **Actual 2018** | Lower bound | **Forecast 2018** | Upper bound | **Relative error** | **Actual 2019** | Lower bound | **Forecast 2019** | Upper bound | **Relative error** |
|---|---|---|---|---|---|---|---|---|---|---|---|
| Austria | 1 | 83975 | 79989 | 83838 | 87687 | 0.2% | 83386 | 80330 | 84778 | 89226 | 1.7% |
| Belgium | 2 | 110693 | 104474 | 110064 | 115654 | 0.6% | 108783 | 104088 | 110467 | 116846 | 1.5% |
| Bulgaria | 3 | 108526 | 104998 | 109178 | 113358 | 0.6% | 108083 | 104614 | 109487 | 114360 | 1.3% |
| Croatia | 4 | 52706 | 50891 | 53598 | 56304 | 1.7% | 51794 | 51026 | 54134 | 57243 | 4.5% |
| Cyprus | 5 | 5768 | 5241 | 5923 | 6606 | 2.7% | 6239 | 5231 | 6030 | 6830 | 3.3% |
| Czechia | 6 | 112920 | 105952 | 111013 | 116075 | 1.7% | 112362 | 105919 | 111762 | 117605 | 0.5% |
| Denmark | 7 | 55232 | 52457 | 54027 | 55597 | 2.2% | 53958 | 44899 | 54592 | 64285 | 1.2% |
| Estonia | 8 | 15751 | 15021 | 15485 | 15949 | 1.7% | 15401 | 14986 | 15563 | 16141 | 1.1% |
| Finland | 9 | 54527 | 53430 | 54309 | 55187 | 0.4% | 53949 | 52814 | 54977 | 57139 | 1.9% |
| France | 10 | 609747 | 589106 | 614377 | 639648 | 0.8% | 613388 | 595850 | 624595 | 653339 | 1.8% |
| Germany | 11 | 954874 | 902512 | 943759 | 985005 | 1.2% | 939520 | 909437 | 956789 | 1004140 | 1.8% |
| Greece | 12 | 120296 | 117941 | 125244 | 132546 | 4.1% | 124965 | 119272 | 127644 | 136016 | 2.1% |
| Hungary | 13 | 131247 | 126031 | 131061 | 136091 | 0.1% | 129892 | 126220 | 132160 | 138100 | 1.7% |
| Ireland | 14 | 31140 | 29127 | 30392 | 31656 | 2.4% | 30959 | 27669 | 30758 | 33847 | 0.6% |
| Italy | 15 | 633133 | 611894 | 650855 | 689816 | 2.8% | 634417 | 615328 | 660375 | 705422 | 4.1% |
| Latvia | 16 | 28820 | 27488 | 28252 | 29015 | 2.0% | 27719 | 27361 | 28125 | 28889 | 1.5% |
| Lithuania | 17 | 39574 | 38861 | 40507 | 42153 | 2.4% | 38281 | 38474 | 40326 | 42177 | 5.3% |
| Luxembourg | 18 | 4318 | 4024 | 4232 | 4440 | 2.0% | 4283 | 4067 | 4310 | 4552 | 0.6% |
| Malta | 19 | 3688 | 3253 | 3552 | 3851 | 3.7% | 3688 | 3258 | 3601 | 3943 | 2.4% |
| Netherlands | 20 | 153363 | 147484 | 152972 | 158460 | 0.3% | 151885 | 149351 | 155515 | 161680 | 2.4% |
| Poland | 21 | 414200 | 383966 | 401787 | 419609 | 3.0% | 409709 | 385809 | 406369 | 426929 | 0.8% |
| Portugal | 22 | 113051 | 106349 | 111178 | 116007 | 1.7% | 111793 | 106749 | 112237 | 117725 | 0.4% |
| Romania | 23 | 264963 | 252485 | 263670 | 274855 | 0.5% | 259889 | 253447 | 265985 | 278524 | 2.3% |
| Slovakia | 24 | 54293 | 51733 | 53945 | 56157 | 0.6% | 53234 | 51863 | 54501 | 57139 | 2.4% |
| Slovenia | 25 | 20485 | 19720 | 20503 | 21286 | 0.1% | 20588 | 19925 | 20813 | 21701 | 1.1% |
| Spain | 26 | 425153 | 404561 | 426577 | 448592 | 0.3% | 416102 | 407553 | 432807 | 458060 | 4.0% |
| Sweden | 27 | 92185 | 88894 | 91592 | 94291 | 0.6% | 88766 | 88936 | 91922 | 94908 | 3.6% |
| | | | | | **MAPE=** | **1.5%** | | | | **MAPE=** | **2.1%** |

*Note*: the red color highlights that the actual value is lower than the lower bound of the 95% confidence interval of the forecasted value.

**Table 3. Predicted DC (by CC and by EM) and officially declared DC, for the 27 EU countries, in 2020 and 2021.**

| Country | Country code | DC 2020 predicted by CC | DC 2020 predicted by EM | Declared DC 2020 | DC 2021 predicted by CC | DC 2021 predicted by EM | Declared DC 2021 |
|---|---|---|---|---|---|---|---|
| Austria | 1 | 6832 | 5930 | 6431 | 9813 | 8212 | 9197 |
| Belgium | 2 | 13686 | 12676 | 19819 | 16743 | 2251 | 8563 |
| Bulgaria | 3 | 3451 | 12726 | 7515 | 5501 | 29311 | 23375 |
| Croatia | 4 | 3625 | 4065 | 3920 | 5047 | 11311 | 8618 |
| Cyprus | 5 | 162 | 377 | 78 | 1246 | 2142 | 568 |
| Czechia | 6 | 15759 | 12738 | 12016 | 20286 | 21716 | 24339 |
| Denmark | 7 | 2550 | 1151* | 907 | 6201 | 1151 | 1938 |
| Estonia | 8 | 340 | 73 | 233 | 1934 | 4968 | 1692 |
| Finland | 9 | 460 | 725 | 598 | 2215 | 4290 | 1150 |
| France | 10 | 70905 | 28966 | 44826 | 100647 | 19555 | 59135 |
| Germany | 11 | 44161 | 14856 | 51010 | 71900 | 31846 | 67134 |
| Greece | 12 | 2214 | 4046 | 4838 | 11715 | 15291 | 20790 |
| Hungary | 13 | 5985 | 8709 | 9537 | 10042 | 21058 | 29649 |
| Ireland | 14 | 1358 | 422 | 2239 | 7240 | 2941 | 3675 |
| Italy | 15 | 54823 | 59057 | 74159 | 51256 | 36872 | 63243 |
| Latvia | 16 | 523 | 895 | 635 | 2158 | 8667 | 3935 |
| Lithuania | 17 | 2387 | 4965 | 1845 | 3665 | 11437 | 5571 |
| Luxembourg | 18 | 529 | 322 | 396 | 453 | 85 | 407 |
| Malta | 19 | 112 | 480 | 166 | 295 | 1186 | 258 |
| Netherlands | 20 | 17369 | 10670 | 11405 | 27964 | 13894 | 9514 |
| Poland | 21 | 30862 | 39849 | 28554 | 34418 | 53543 | 68500 |
| Portugal | 22 | 8029 | 8625 | 6906 | 11007 | 11869 | 11989 |
| Romania | 23 | 13245 | 23164 | 15767 | 12999 | 41956 | 42985 |
| Slovakia | 24 | 5175 | 4812 | 2250 | 11936 | 17845 | 14415 |
| Slovenia | 25 | 1938 | 3171 | 3025 | 3267 | 3999 | 2885 |
| Spain | 26 | 51833 | 39567 | 54374 | 62706 | 14866 | 36850 |
| Sweden | 27 | 8978 | 6467 | 9817 | 9419 | 2506 | 5534 |

*Note*: *Estimation based on natural scale regression.

**Table 4. Standardized values of min, max predicted and declared DC. Discrepancy index $I_D$. Pandemic fatal impact index $I_F$.**

| Country code | P | A | d1 (2020) | d2 (2020) | d3 (2020) | $I_D$ (2020) | d1 (2021) | d2 (2021) | d3 (2021) | $I_D$ (2021) | $\overline{DC}$ (2020) | $I_F$ (2020) | $\overline{DC}$ (2021) | $I_F$ (2021) |
|---|---|---|---|---|---|---|---|---|---|---|---|---|---|---|
| 1 | 8858775 | 82409 | 1.0E-03 | 1.2E-03 | 1.1E-03 | 1.6E-04 | 1.4E-03 | 1.7E-03 | 1.6E-03 | 3.1E-04 | 6381 | 8.74E-09 | 9012 | 1.23E-08 |
| 2 | 11455519 | 30280 | 2.3E-03 | 2.5E-03 | 3.6E-03 | 1.4E-03 | 4.1E-04 | 3.0E-03 | 1.6E-03 | 2.8E-03 | 13181 | 3.80E-08 | 9497 | 2.74E-08 |
| 3 | 7000039 | 108560 | 2.6E-03 | 9.6E-03 | 5.7E-03 | 7.4E-03 | 4.2E-03 | 2.2E-02 | 1.8E-02 | 2.2E-02 | 8088 | 1.06E-08 | 17406 | 2.29E-08 |
| 4 | 4076246 | 55960 | 4.4E-05 | 4.9E-05 | 4.7E-05 | 6.2E-06 | 6.1E-05 | 1.4E-04 | 1.0E-04 | 8.1E-05 | 3845 | 1.69E-08 | 8179 | 3.59E-08 |
| 5 | 875899 | 9240 | 3.3E-05 | 7.7E-05 | 1.6E-05 | 8.3E-05 | 2.5E-04 | 4.4E-04 | 1.2E-04 | 4.1E-04 | 269 | 3.33E-08 | 1694 | 2.09E-07 |
| 6 | 10649800 | 77240 | 1.9E-04 | 2.4E-04 | 1.8E-04 | 7.8E-05 | 3.0E-04 | 3.2E-04 | 3.6E-04 | 7.1E-05 | 14248 | 1.73E-08 | 21001 | 2.55E-08 |
| 7 | 5806081 | 42430 | 1.1E-04 | 2.5E-04 | 8.9E-05 | 2.3E-04 | 1.1E-04 | 6.1E-04 | 1.9E-04 | 6.6E-04 | 1851 | 7.51E-09 | 3676 | 1.49E-08 |
| 8 | 1324820 | 42390 | 1.6E-06 | 7.2E-06 | 5.0E-06 | 6.2E-06 | 4.1E-05 | 1.1E-04 | 3.6E-05 | 1.0E-04 | 207 | 3.68E-09 | 3451 | 6.14E-08 |
| 9 | 5517919 | 303890 | 4.3E-05 | 6.8E-05 | 5.6E-05 | 2.5E-05 | 2.1E-04 | 4.0E-04 | 1.1E-04 | 3.9E-04 | 593 | 3.53E-10 | 3253 | 1.94E-09 |
| 10 | 67012883 | 547557 | 7.6E-04 | 1.9E-03 | 1.2E-03 | 1.2E-03 | 5.1E-04 | 2.7E-03 | 1.6E-03 | 2.2E-03 | 49936 | 1.36E-09 | 60101 | 1.64E-09 |
| 11 | 83019213 | 348560 | 2.5E-04 | 7.3E-04 | 8.5E-04 | 8.4E-04 | 5.3E-04 | 1.2E-03 | 1.1E-03 | 9.2E-04 | 29509 | 1.02E-09 | 51873 | 1.79E-09 |
| 12 | 10724599 | 128900 | 1.2E-03 | 2.1E-03 | 2.5E-03 | 1.8E-03 | 6.1E-03 | 8.0E-03 | 1.1E-02 | 5.7E-03 | 3130 | 2.26E-09 | 13503 | 9.77E-09 |
| 13 | 9772756 | 90530 | 3.1E-04 | 4.5E-04 | 4.9E-04 | 2.5E-04 | 5.2E-04 | 1.1E-03 | 1.5E-03 | 1.3E-03 | 7347 | 8.30E-09 | 15550 | 1.76E-08 |
| 14 | 4904240 | 68890 | 4.8E-05 | 1.5E-04 | 2.5E-04 | 2.6E-04 | 3.3E-04 | 8.2E-04 | 4.1E-04 | 6.5E-04 | 890 | 2.64E-09 | 5090 | 1.51E-08 |
| 15 | 60359546 | 294140 | 5.3E-03 | 5.7E-03 | 7.2E-03 | 2.1E-03 | 3.6E-03 | 5.0E-03 | 6.2E-03 | 3.3E-03 | 56940 | 3.21E-09 | 44064 | 2.48E-09 |
| 16 | 1919968 | 62200 | 5.4E-05 | 9.2E-05 | 6.5E-05 | 4.6E-05 | 2.2E-04 | 8.9E-04 | 4.0E-04 | 8.2E-04 | 709 | 5.94E-09 | 5413 | 4.53E-08 |
| 17 | 2794184 | 62674 | 5.9E-04 | 1.2E-03 | 4.5E-04 | 1.1E-03 | 9.0E-04 | 2.8E-03 | 1.4E-03 | 2.4E-03 | 3676 | 2.10E-08 | 7551 | 4.31E-08 |
| 18 | 613894 | 2590 | 5.9E-05 | 9.7E-05 | 7.3E-05 | 4.3E-05 | 1.6E-05 | 8.3E-05 | 7.5E-05 | 9.3E-05 | 426 | 2.68E-07 | 269 | 1.69E-07 |
| 19 | 493559 | 320 | 1.1E-05 | 4.5E-05 | 1.6E-05 | 4.7E-05 | 2.8E-05 | 1.1E-04 | 2.4E-05 | 1.3E-04 | 296 | 1.88E-06 | 741 | 4.69E-06 |
| 20 | 17282163 | 33720 | 1.5E-03 | 2.5E-03 | 1.6E-03 | 1.3E-03 | 2.0E-03 | 4.0E-03 | 1.4E-03 | 3.6E-03 | 14020 | 2.41E-08 | 20929 | 3.59E-08 |
| 21 | 37972812 | 306230 | 1.8E-03 | 2.3E-03 | 1.7E-03 | 9.1E-04 | 2.0E-03 | 3.1E-03 | 4.0E-03 | 2.5E-03 | 35356 | 3.04E-09 | 43980 | 3.78E-09 |
| 22 | 10276617 | 91590 | 2.9E-03 | 3.1E-03 | 2.5E-03 | 7.2E-04 | 3.9E-03 | 4.2E-03 | 4.3E-03 | 5.1E-04 | 8327 | 8.85E-09 | 11438 | 1.22E-08 |
| 23 | 19414458 | 230170 | 1.5E-02 | 2.6E-02 | 1.8E-02 | 1.4E-02 | 1.5E-02 | 4.8E-02 | 4.9E-02 | 5.1E-02 | 18205 | 4.07E-09 | 27477 | 6.15E-09 |
| 24 | 5450421 | 48088 | 4.2E-04 | 4.5E-04 | 2.0E-04 | 2.7E-04 | 1.0E-03 | 1.6E-03 | 1.3E-03 | 5.6E-04 | 4993 | 1.91E-08 | 14891 | 5.68E-08 |
| 25 | 2080908 | 20140 | 9.3E-04 | 1.5E-03 | 1.5E-03 | 8.2E-04 | 1.6E-03 | 1.9E-03 | 1.4E-03 | 7.1E-04 | 2555 | 6.10E-08 | 3633 | 8.67E-08 |
| 26 | 46937060 | 498800 | 6.4E-02 | 8.4E-02 | 8.9E-02 | 3.4E-02 | 2.4E-02 | 1.0E-01 | 6.0E-02 | 8.1E-02 | 45700 | 1.95E-09 | 38786 | 1.66E-09 |
| 27 | 10230185 | 410340 | 1.3E-02 | 1.8E-02 | 2.0E-02 | 9.3E-03 | 5.1E-03 | 1.9E-02 | 1.1E-02 | 1.5E-02 | 7723 | 1.84E-09 | 5962 | 1.42E-09 |

*Notes*: P=Population; d1 = (std) min predicted DC; d2=(std) max predicted DC; d3=(std) declared DC; A= Land Area [$km^2$]; $\overline{DC}$ = Mean Predicted DC.

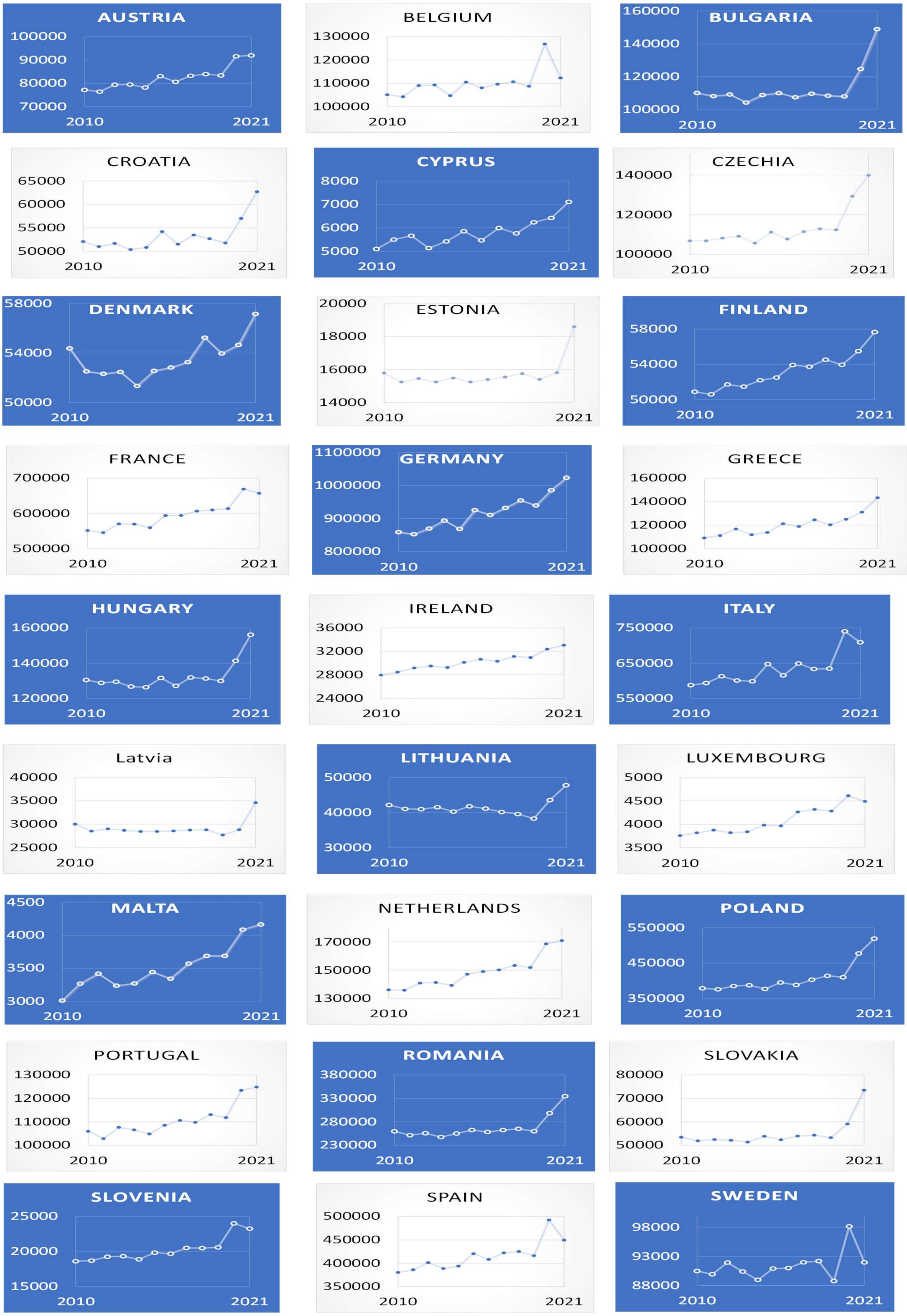

*Figure 1. All-cause mortality trends from 2010--2021 in the 27 EU countries (alphabetic order).*

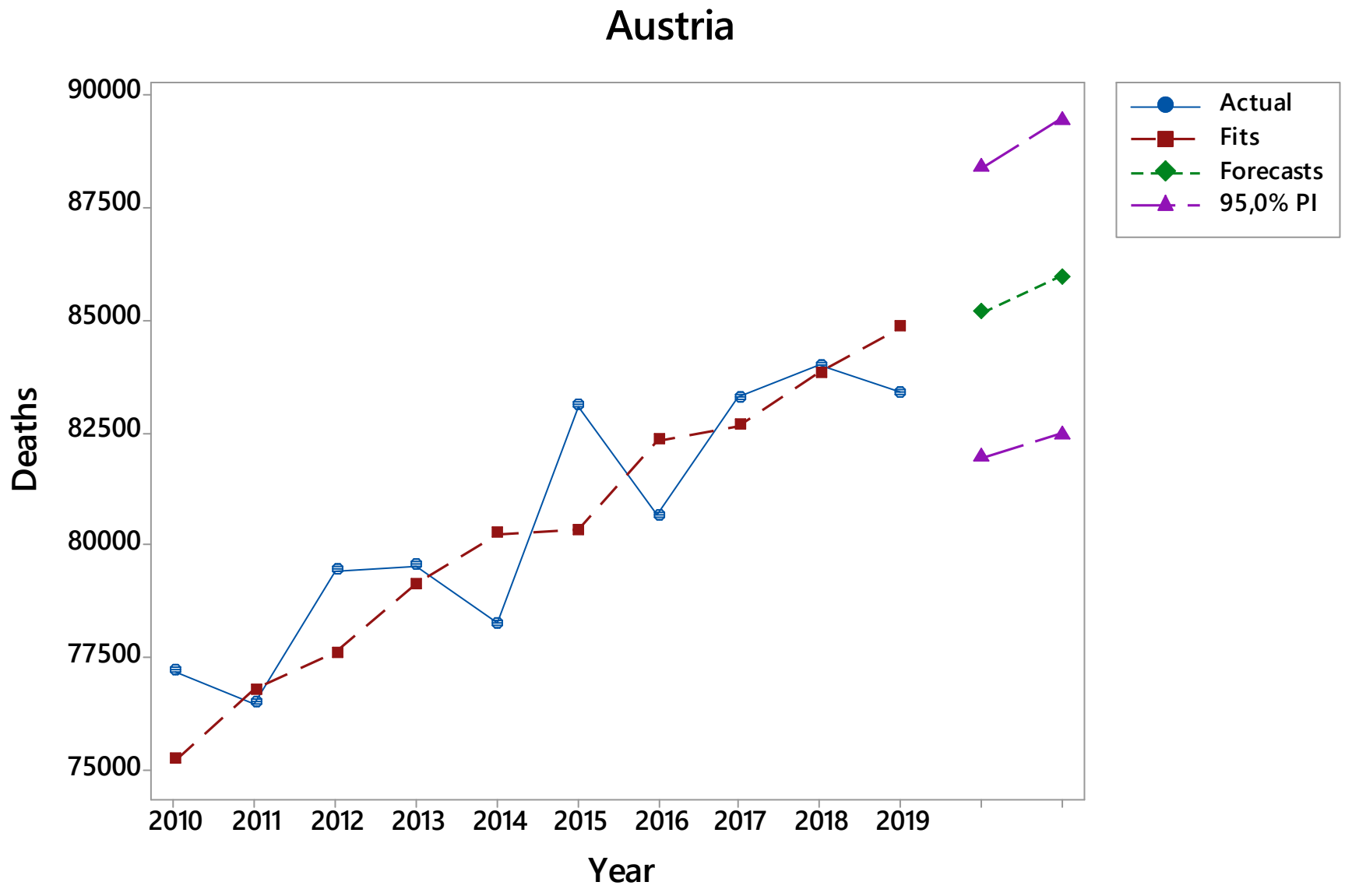

Smoothing constants

$\alpha = 0.32$; $\gamma = 0.41$

| Year | Forecast | LB | UB |
|---|---|---|---|
| 2020 | 85153.1 | 81942.7 | 88363.5 |
| 2021 | 85937.9 | 82442.6 | 89433.3 |

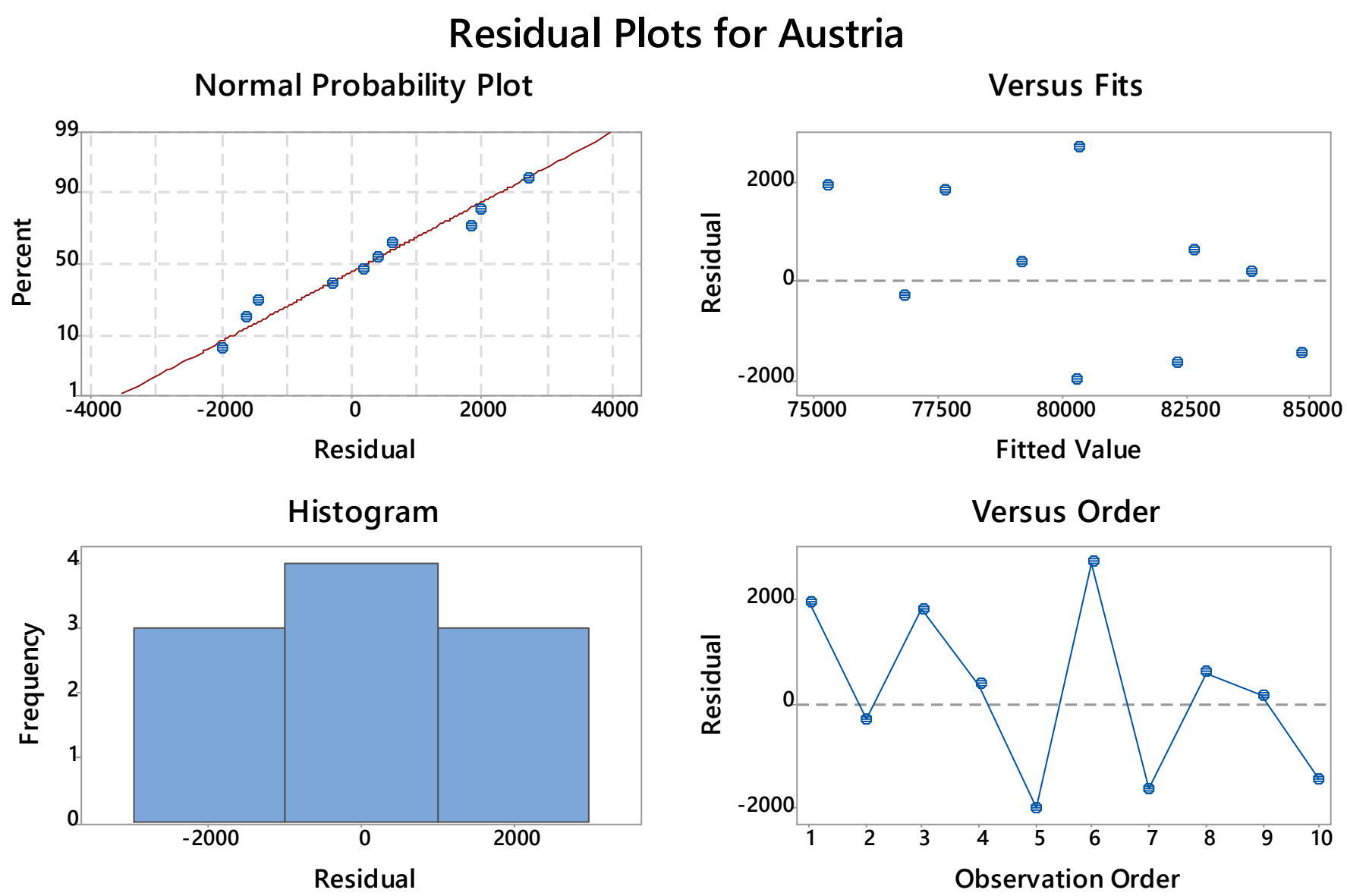

***Figure 2. Example of double exponential smoothing of the (yearly) all-cause mortality time series 2010--2019 in Austria and forecasting for 2020 and 2021.***

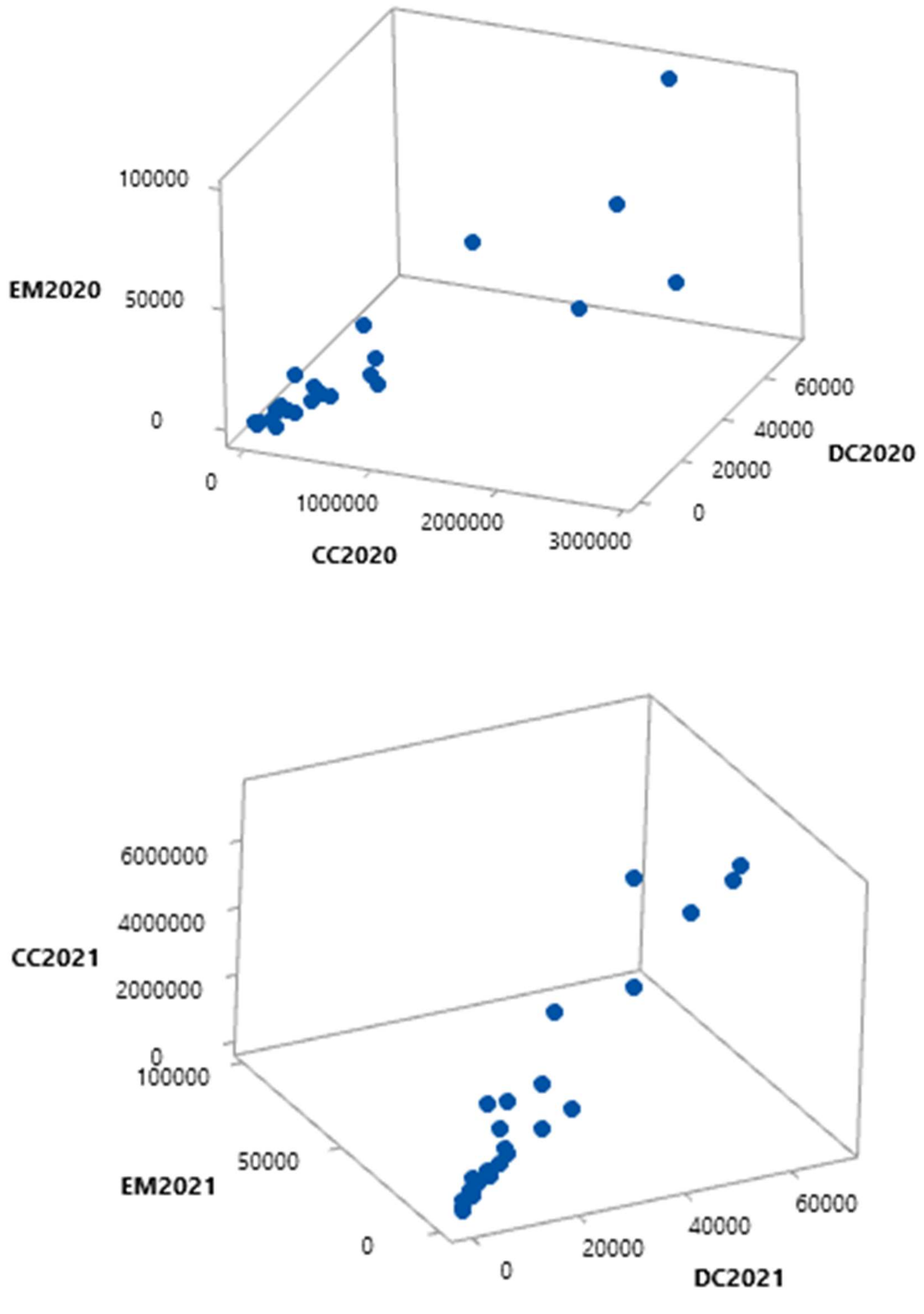

*Figure 3. Scatterplots 3D of the variables Excess Mortality (EM), Covid Cases (CC), Deaths by Covid (DC) for the 27 EU countries in 2020 and 2021.*

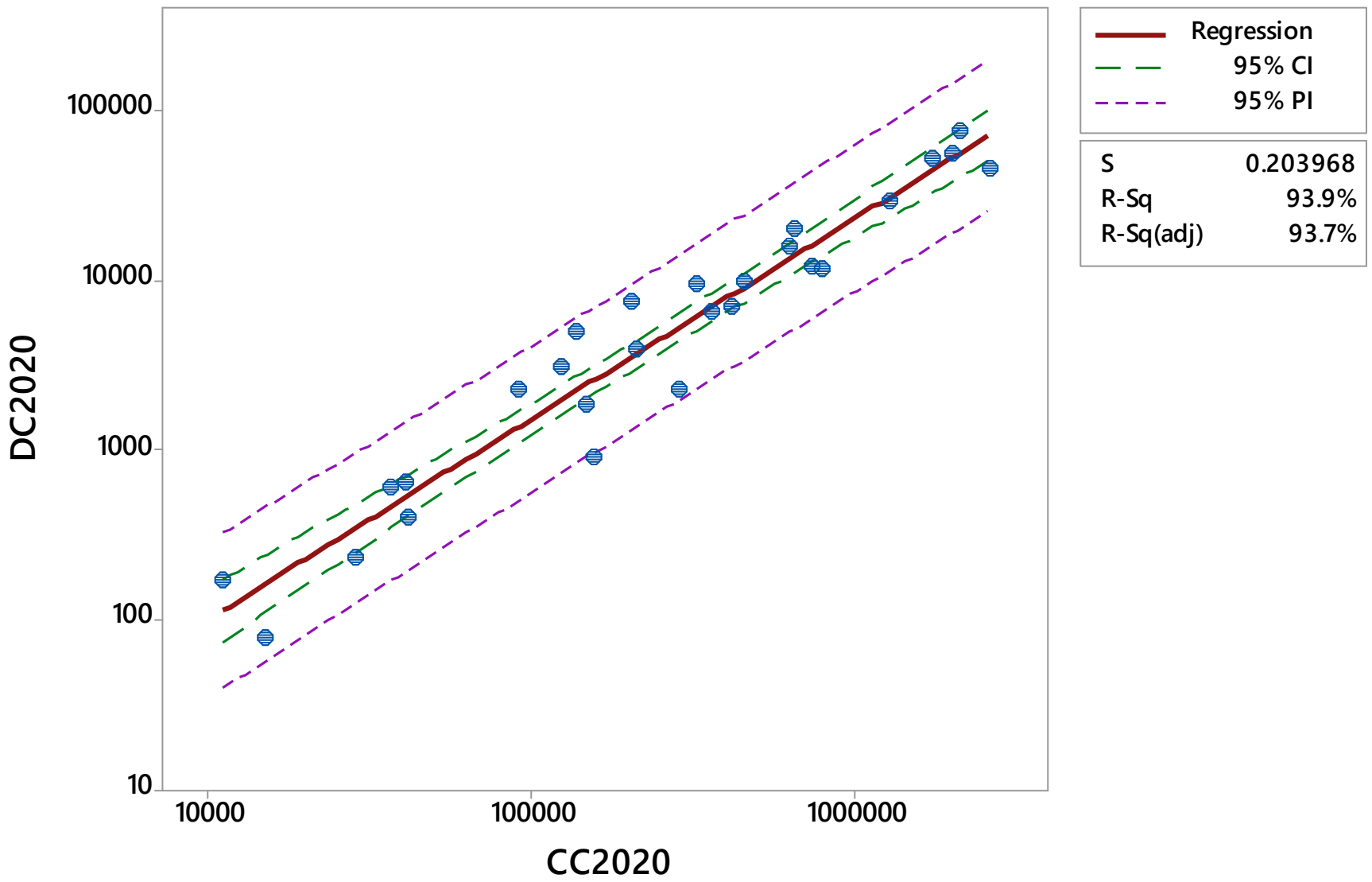

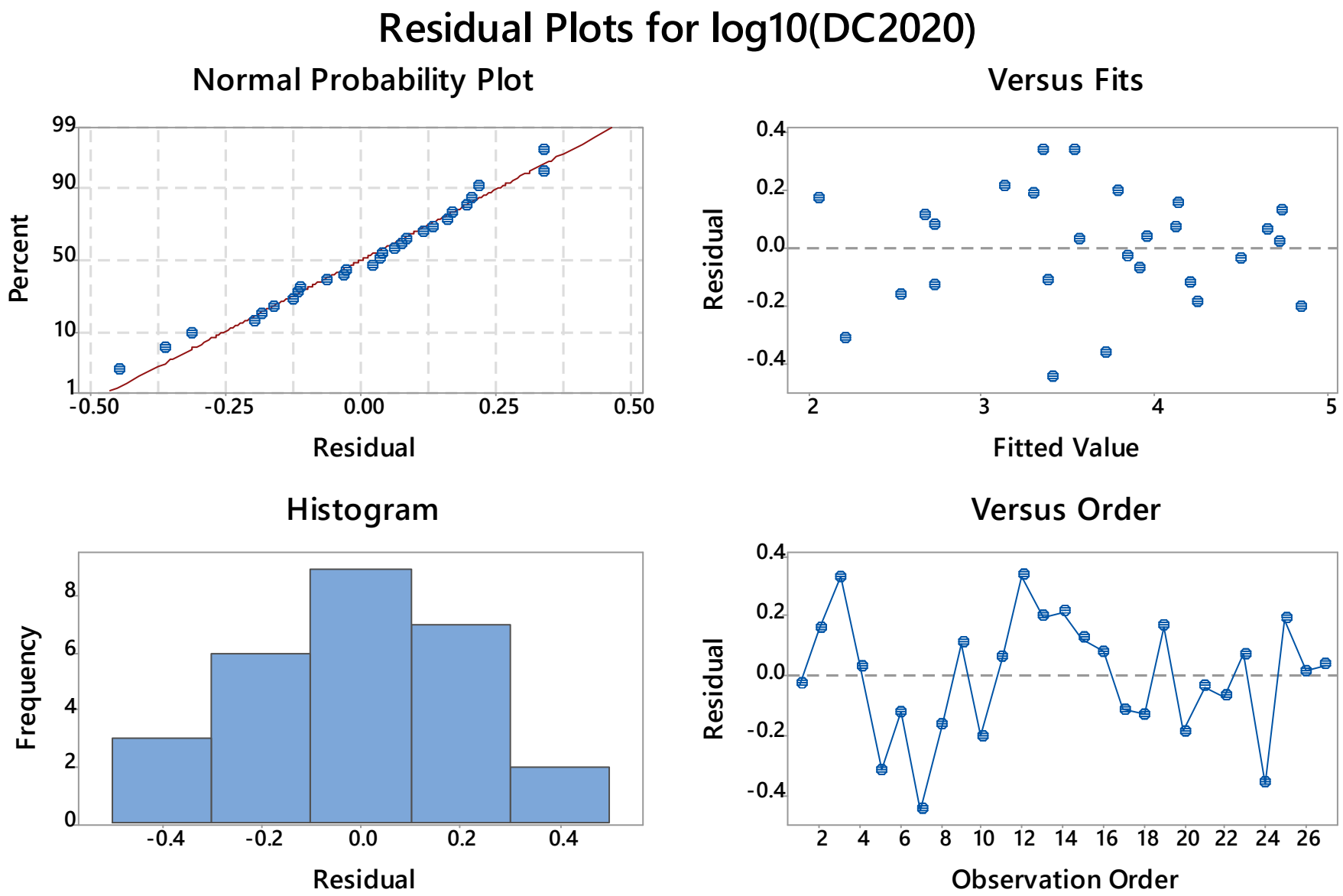

**Estimated regression equation:**

$$\text{Log10(DC2020)} = -2.723 + 1.180\ \text{Log10(CC2020)}$$

**Analysis of Variance**

| *Source* | DF | SS | MS | F ratio | p value |
|---|---|---|---|---|---|
| *Regression* | 1 | 15.99 | 15.99 | 384.45 | 0.000 |
| *Error* | 25 | 1.04 | 0.04 | | |
| *Total* | 26 | 17.03 | | | |

***Figure 4. Regression (log10 transformed) DC vs. CC for the EU countries in 2020.***

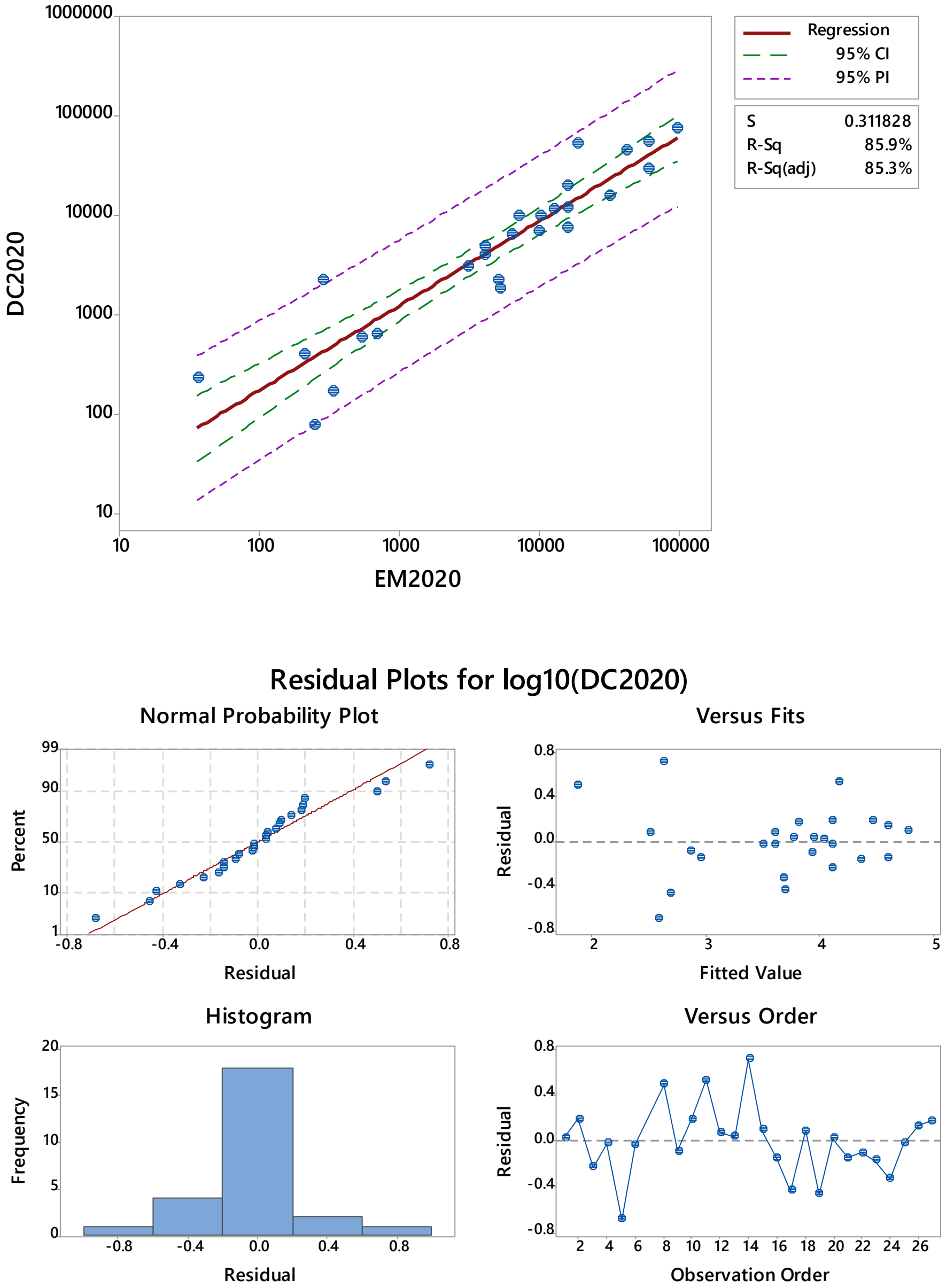

## Estimated regression equation:

Log10(DC2020) = 0.5538 + 0.8451 Log10(EM2020)

## Analysis of Variance

| *Source* | DF | SS | MS | F ratio | p value |
|---|---|---|---|---|---|
| *Regression* | 1 | 14.24 | 14.24 | 146.42 | 0.000 |
| *Error* | 24 | 2.33 | 0.10 | | |
| *Total* | 25 | 16.57 | | | |

***Figure 5. Regression (log10 transformed) DC vs. EM for the EU countries in 2020***

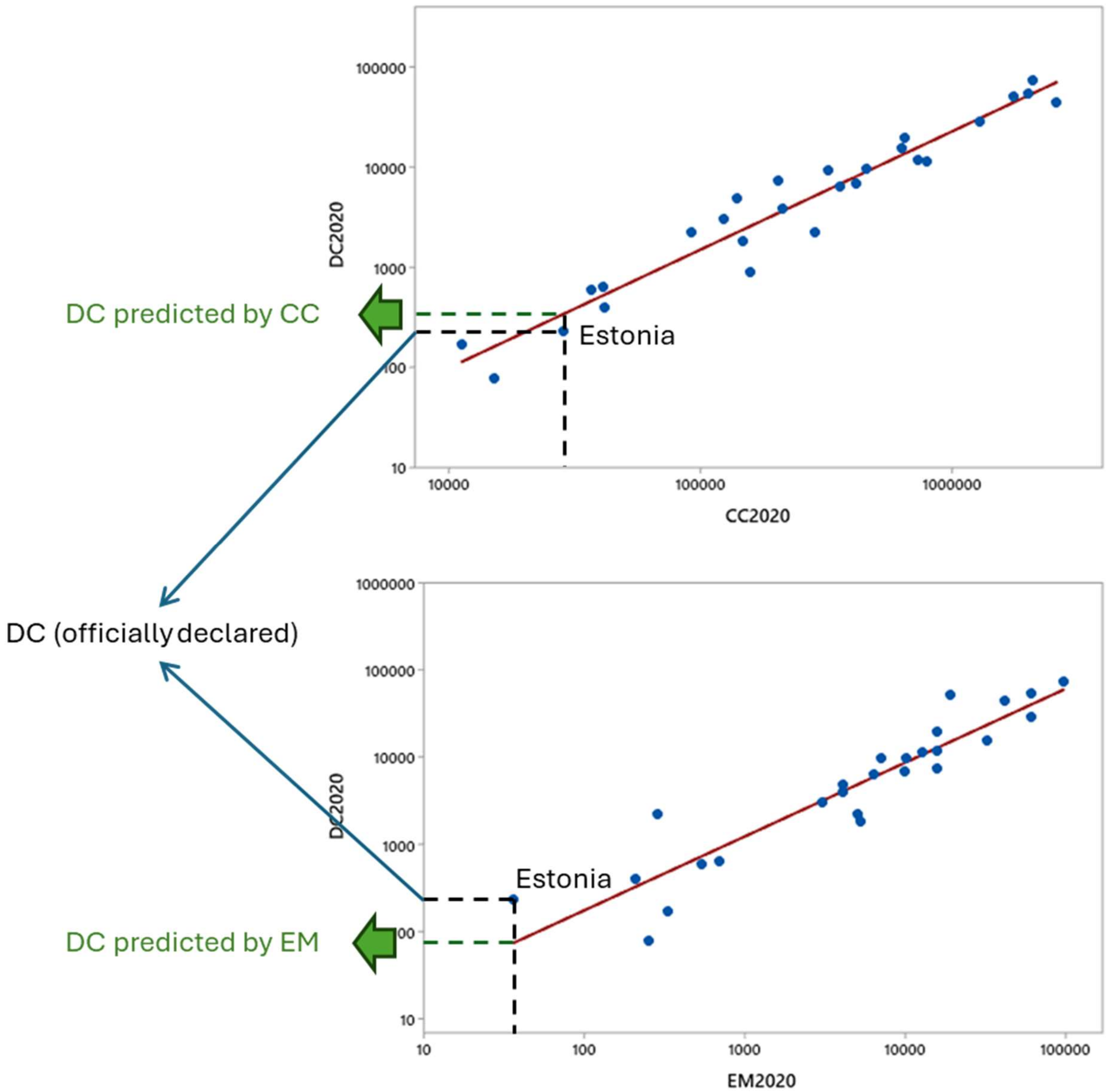

***Figure 6. Scheme of the estimation strategy of COVID-19 mortality (the depicted example concerns Estonia).***

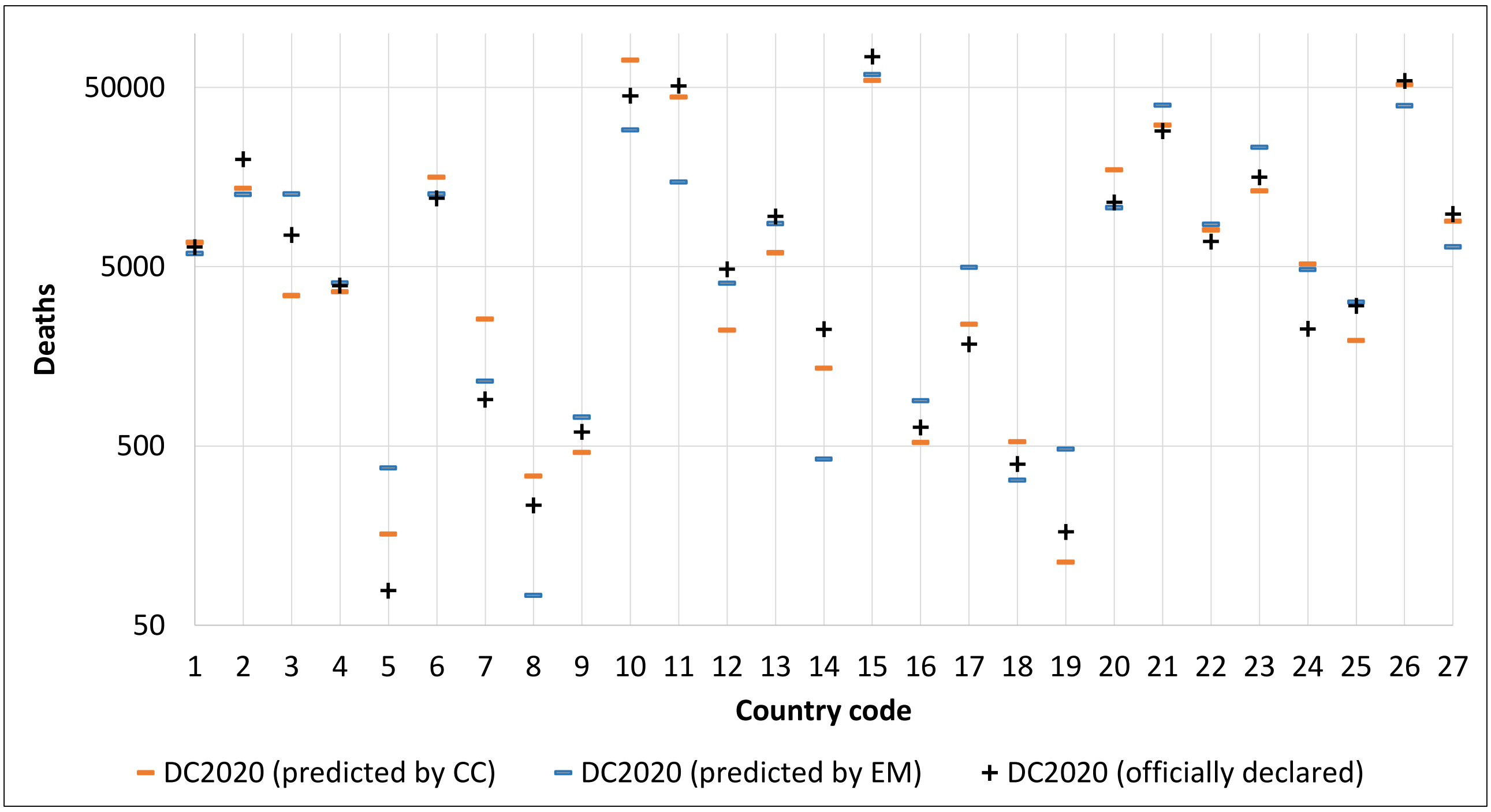

(a)

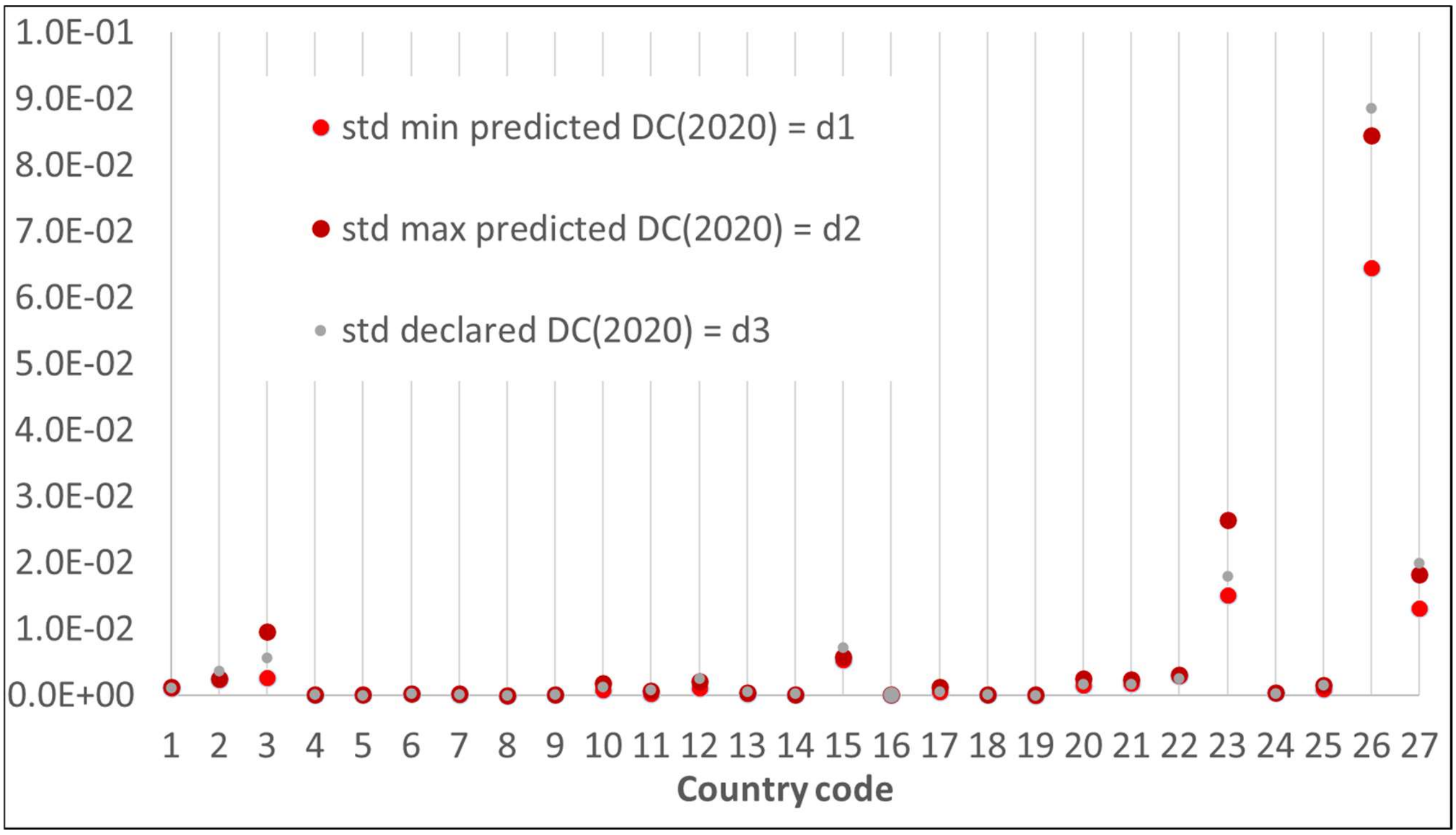

(b)

***Figure 7. For the 27 EU countries in 2020: (a) predicted and declared DC (vertical axis in log scale). (b) Standardized min predicted, max predicted and declared DC.***

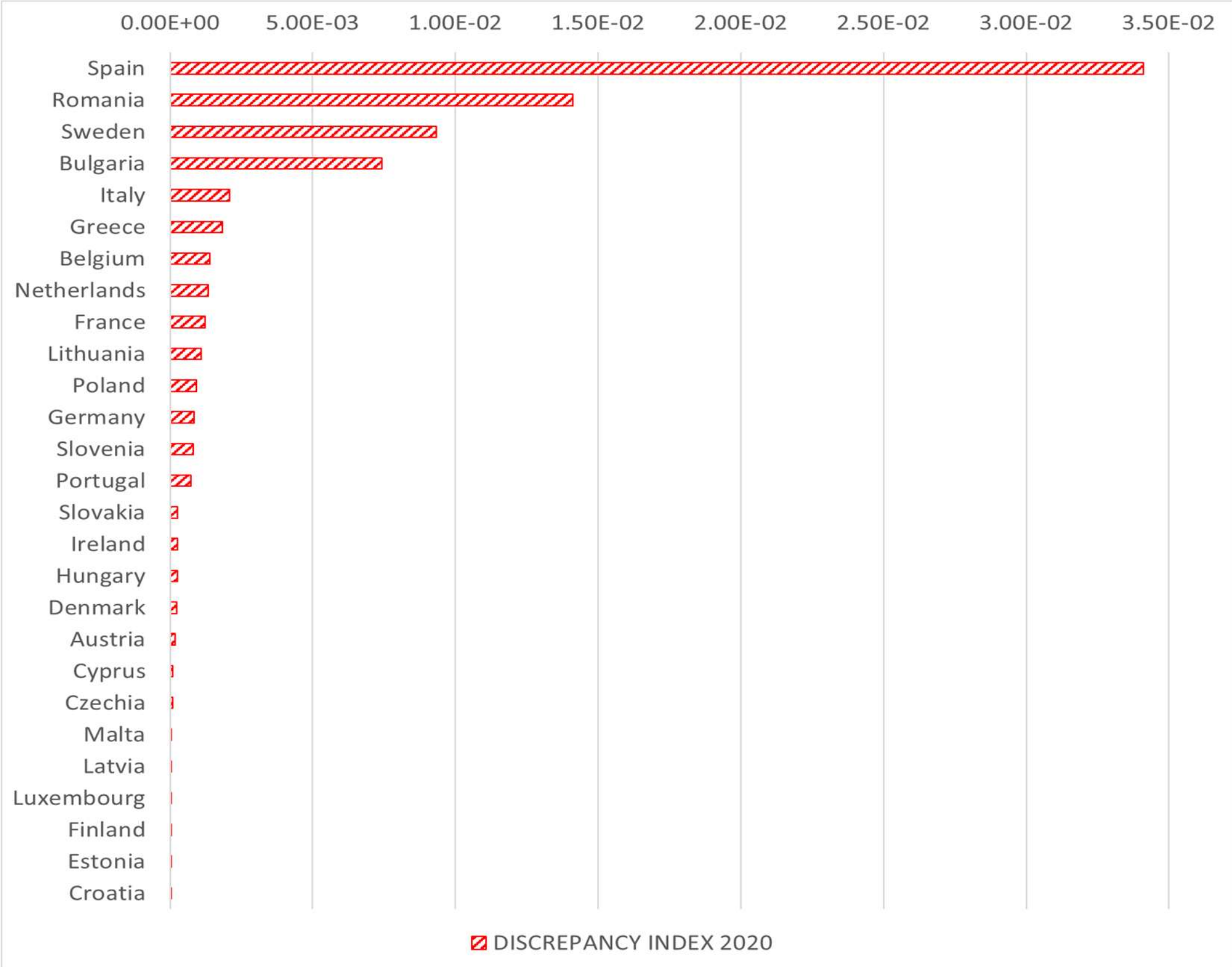

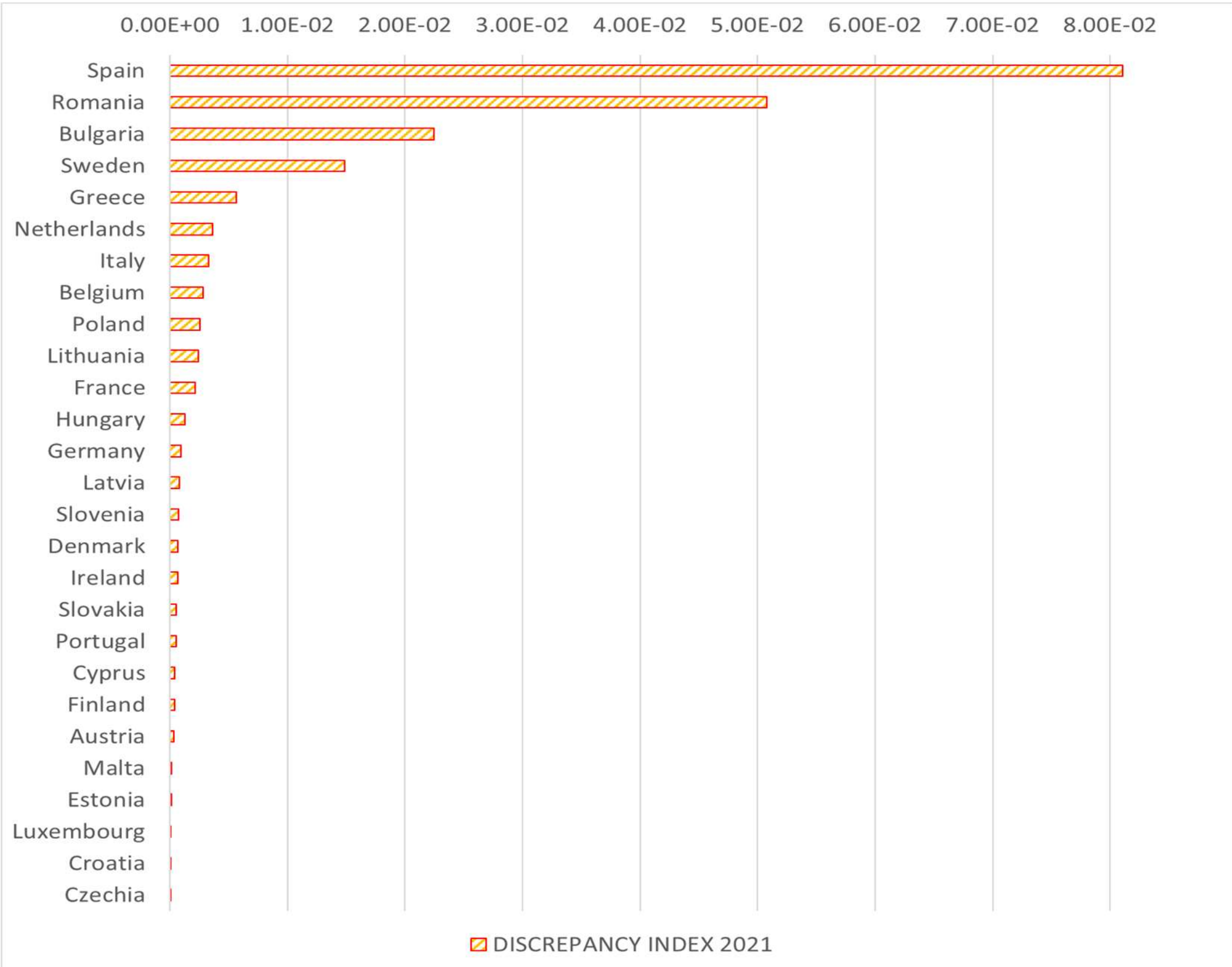

*Figure 8. Discrepancy indices for 2020 and 2021.*

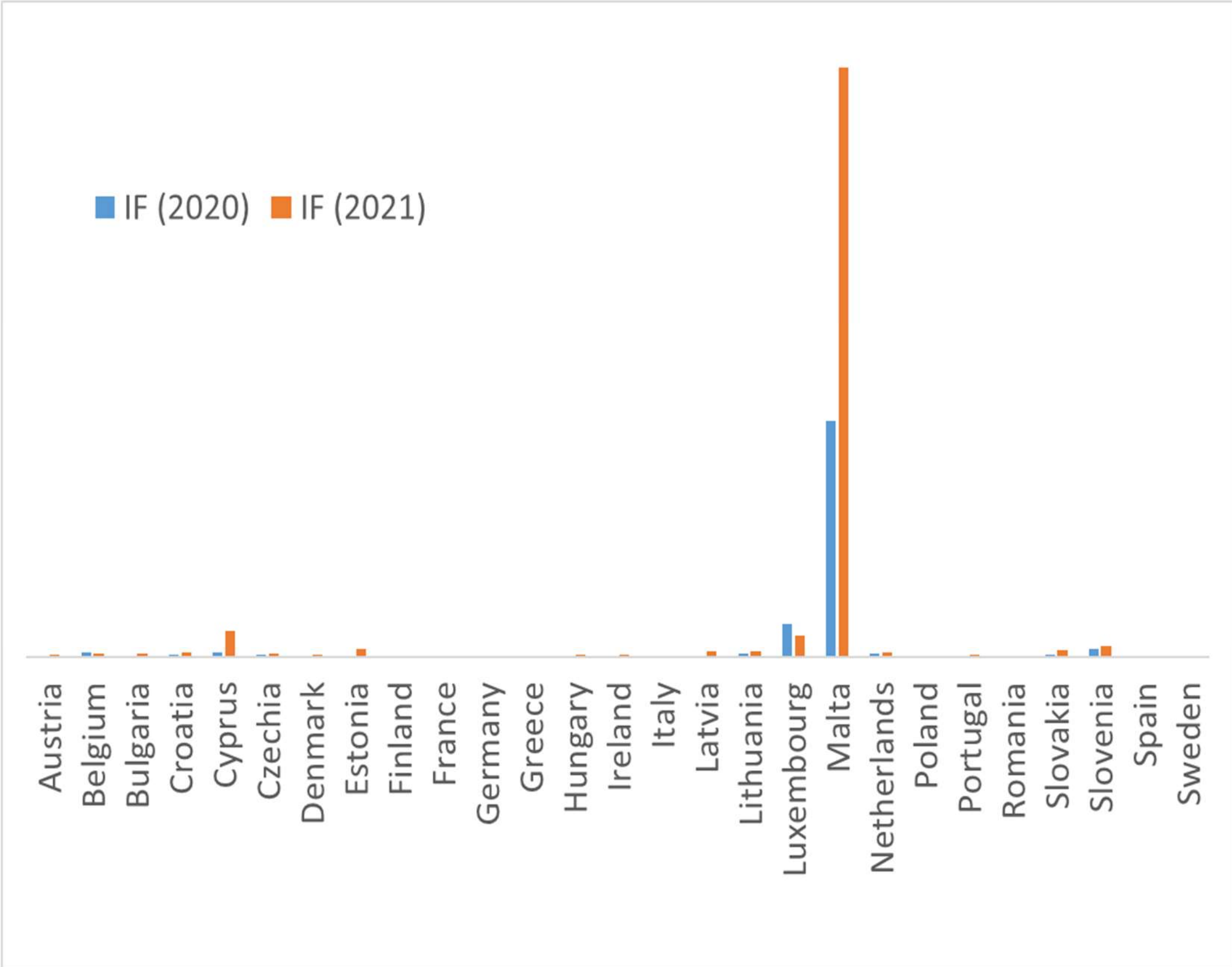

(a)

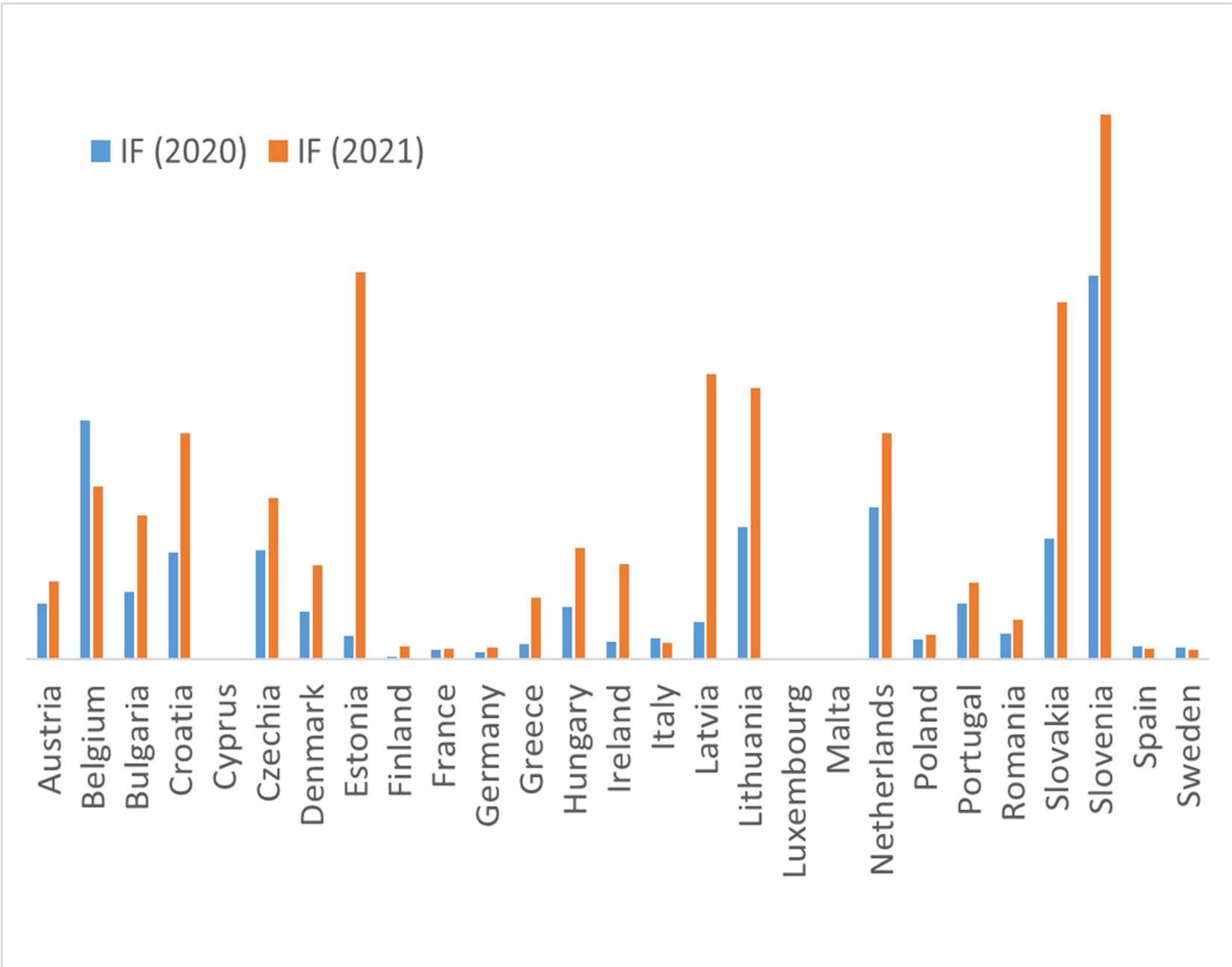

(b)

***Figure 9. (a) Fatal impact index in 2020 and 2021. (b) Rescaled plot without Cyprus, Luxembourg, Malta.***

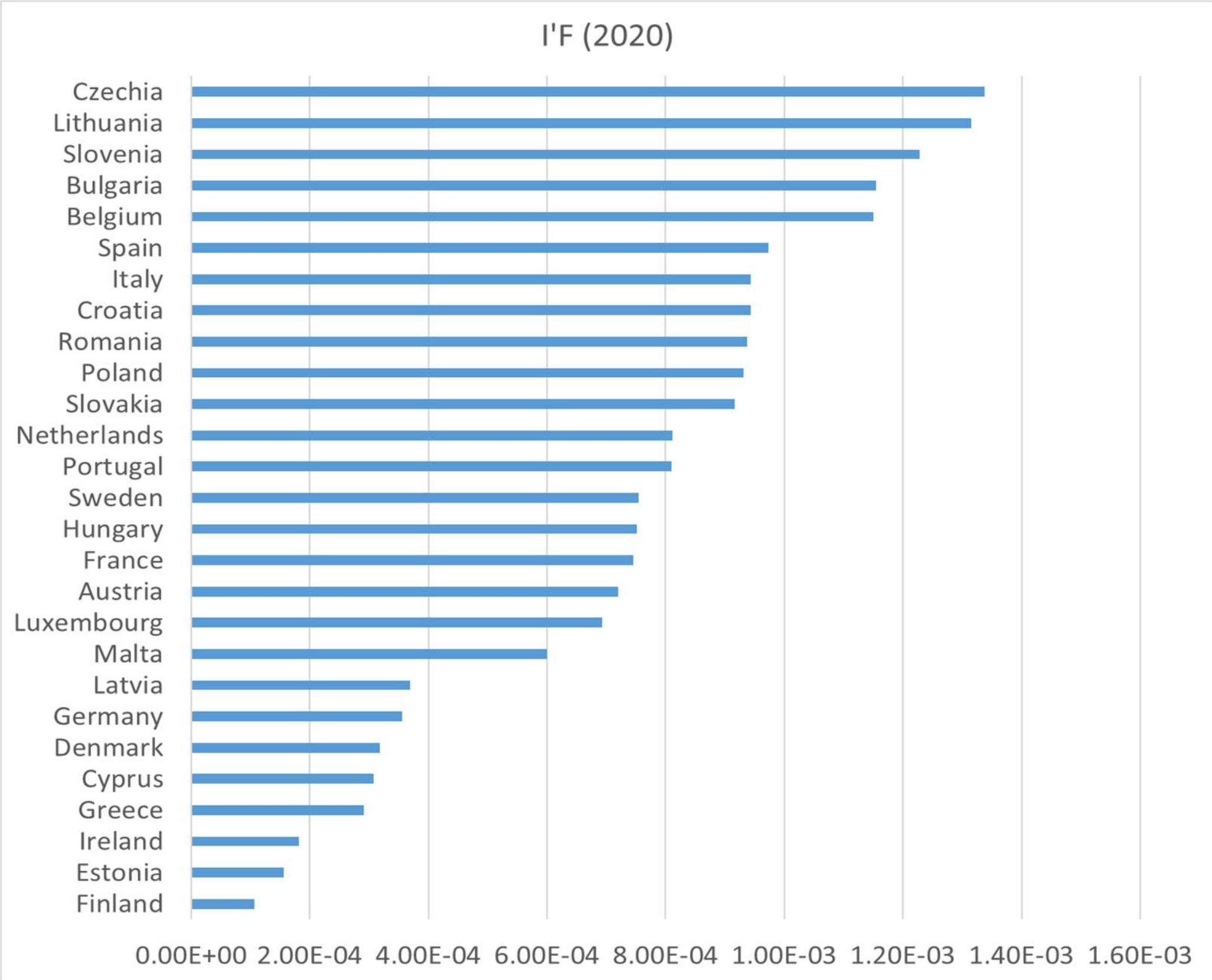

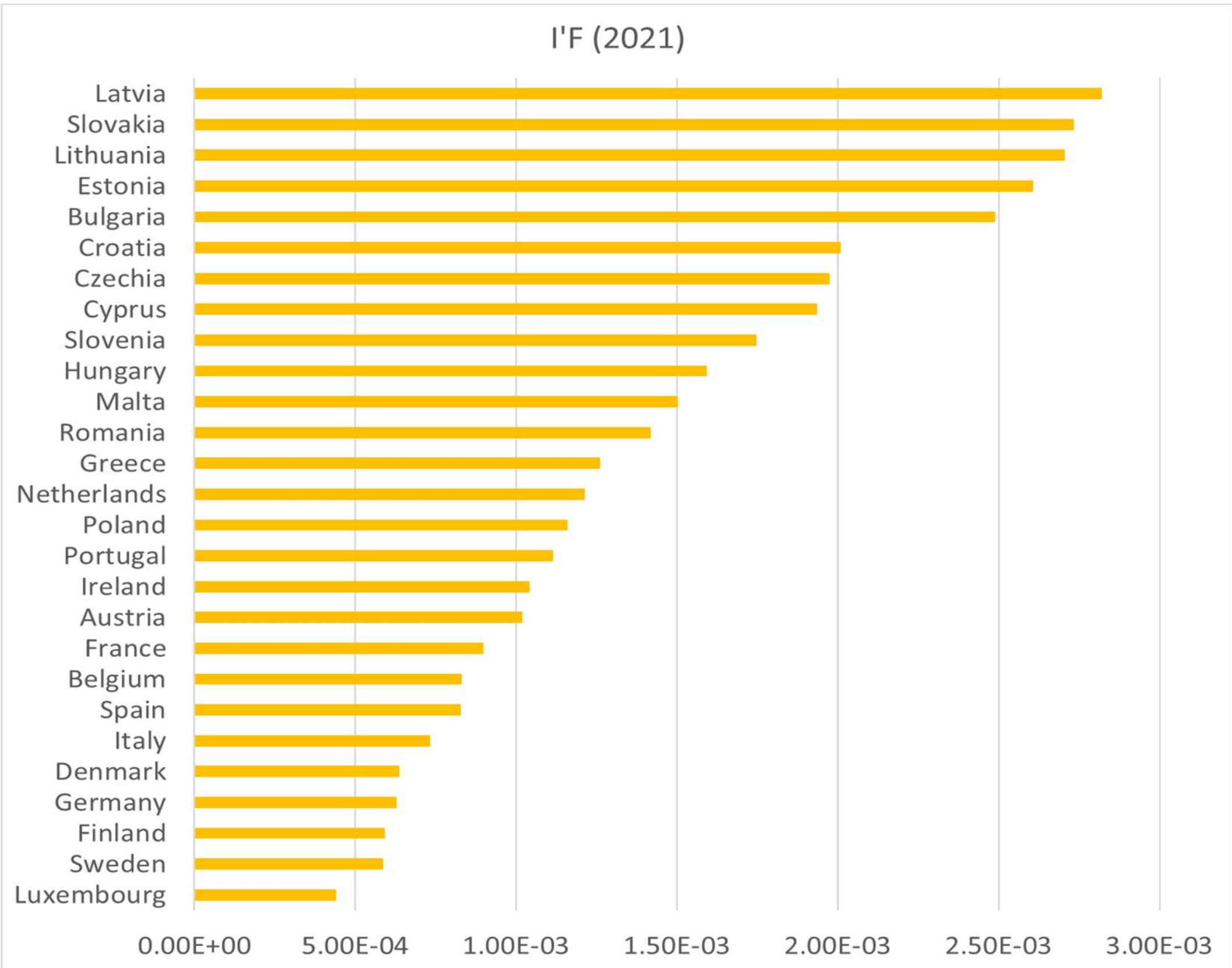

*Figure 10. COVID-19 fatality rates in 2020 and 2021.*